\documentclass[journal]{IEEEtran}

\usepackage{cite}
\usepackage{amsmath,amssymb,amsfonts}
\usepackage{stmaryrd}
\usepackage{algorithm}
\usepackage{algorithmic}
\usepackage{graphicx}
\usepackage{textcomp}
\usepackage{xcolor}
\newif\iftrackedchanges
\trackedchangesfalse
\iftrackedchanges\else
    \colorlet{blue}{black}
\fi
\usepackage{booktabs}
\usepackage{multirow}
\usepackage{subcaption}
\usepackage{hyperref}
\usepackage{url}
\usepackage{siunitx}
\usepackage{float}
\usepackage{placeins}
\usepackage{pgfplots}
\pgfplotsset{compat=1.18}

\newcommand{\appref}[1]{\hyperref[#1]{Appendix~\ref*{#1}}}

\begin{document}

\title{PRICE: Privacy-Preserving Image Classification in Mobile Edge-Cloud Computing with Homomorphic Encryption}

\author{
\IEEEauthorblockN{
Duc-Huy Nguyen, Bui Duc Manh, Chi-Hieu Nguyen, Dinh Thai Hoang\\
Diep N. Nguyen, Van-Dinh Nguyen, and Quoc-Viet Pham
}

\thanks{Duc-Huy Nguyen, Bui Duc Manh, Chi-Hieu Nguyen, Dinh Thai Hoang, and Diep N. Nguyen are with the School of Electrical and Data Engineering, University of Technology Sydney, Sydney, NSW 2007, Australia (e-mail: duc.h.nguyen-3@student.uts.edu.au; duc.m.bui-1@student.uts.edu.au; hieu.c.nguyen@student.uts.edu.au; hoang.dinh@uts.edu.au; diep.nguyen@uts.edu.au).}
\thanks{Van-Dinh Nguyen and Quoc-Viet Pham are with the School of Computer Science and Statistics, Trinity College Dublin, Dublin 2, D02PN40, Ireland (e-mail: dinh.nguyen@tcd.ie; viet.pham@tcd.ie).}
\thanks{Some results of this work were presented at the ACM MobiCom Conference (Demo Paper), 2025~\cite{nguyen2025ppaicloud}.
}
}

\maketitle

\begin{abstract}
Cloud-assisted image classification enables many mobile AI applications, but transmitting raw images to the cloud raises significant privacy concerns. Homomorphic encryption (HE) allows computation over encrypted data; however, existing end-to-end HE-based inference pipelines incur prohibitive latency and computational overhead, limiting their feasibility for latency-sensitive mobile systems. This paper addresses how to enable efficient encrypted image classification in mobile edge-cloud environments without sacrificing accuracy. We propose PRICE, a three-tier edge-cloud architecture that partitions deep neural network inference between a trusted edge and an untrusted cloud. The edge performs lightweight feature extraction in plaintext, while only compact embeddings are encrypted using the Cheon-Kim-Kim-Song (CKKS) scheme and offloaded to the cloud for encrypted final-layer inference. This design avoids homomorphic evaluation of deep backbone layers, significantly reducing computation depth, ciphertext size, and end-to-end latency. To further improve efficiency, we develop RowHyDiag, a row-packed hybrid diagonal encoding with hoisted-rotation-aware encrypted matrix-vector computation, which minimizes rotation operations and improves slot utilization. Extensive experiments on real devices \textcolor{blue}{across five datasets} demonstrate that\textcolor{blue}{, under the evaluated 128-bit-security configurations, RowHyDiag achieves up to 31.8$\times$ faster cloud inference, 85.3$\times$ fewer rotations, and 83.0$\times$ lower setup payload than the conventional Halevi-Shoup diagonal method}. \textcolor{blue}{RowHyDiag also outperforms the PEGASUS linear-transformation baseline, providing 1.1-4.5$\times$ faster cloud inference and 1.2-3.4$\times$ lower setup payload.} The results show that PRICE enables deployment-ready, low-latency encrypted inference for mobile edge-cloud systems under a trusted-edge threat model.
\end{abstract}
\begin{IEEEkeywords}
Encrypted inference, homomorphic encryption, mobile edge-cloud computing, privacy-preserving image classification, low-latency edge-cloud deployment.
\end{IEEEkeywords}

\section{Introduction}
\label{sec:introduction}
\subsection{Motivation}

The rapid advancement of cloud computing has transformed image classification from a primarily on-device task into an on-demand, large-scale service~\cite{Huang_2022,Gao_2023}. As deep learning (DL) models, such as ResNet and EfficientNet, continue to improve visual-recognition accuracy~\cite{he2016resnet,tan2021efficientnetv2}, cloud platforms have emerged as a widely adopted solution for
executing neural network inference for visual recognition. By providing scalable resources and consistent performance, cloud-based inference enables heterogeneous mobile devices to
access powerful models without local execution constraints ~\cite{Liu_2022,Li_2023}. This offloading trend is particularly important for resource-constrained devices (i.e., smartphones and tablets), where running modern DL classifiers locally can be computationally expensive, resource-intensive, and latency-sensitive~\cite{kang2017neurosurgeon, Xu_2021}. Consequently, cloud-assisted image classification has become widely adopted, enabling mobile users to perform interactive visual recognition by leveraging remote computing resources~\cite{Huang_2022,Li_2024}.

Despite these advantages, cloud-based inference introduces a fundamental privacy challenge: users must transmit visual data to external servers that are often modeled as honest-but-curious~\cite{goldreich2004foc2}, creating the risk of exposing sensitive content, such as personal identity, location, and behavioral patterns~\cite{rigaki2023privacy,zhao2025visualprivacy, Wang_2024}. To address this issue, secure multi-party computation (SMPC) has emerged as a representative approach for privacy-preserving DL inference~\cite{zeng2025towards}. In this setting, a private input is split into random shares and distributed to multiple parties, ensuring that each party sees only an unintelligible share while jointly evaluating inference through interactive protocols~\cite{juvekar2018gazelle, wagh2019securenn}. Foundational studies in this domain include Gazelle, a two‑party secure inference framework that combines homomorphic encryption with garbled circuits to achieve 20-30× faster inference than earlier approaches~\cite{juvekar2018gazelle}, and SecureNN, a three‑party protocol that enables efficient convolutional neural network (CNN) training and inference with security guarantees against both semi‑honest and malicious adversaries~\cite{wagh2019securenn}. These studies demonstrate the feasibility of CNN inference under representative SMPC settings with different participant roles and adversary models~\cite{juvekar2018gazelle,wagh2019securenn}.

However, a key limitation of SMPC is the massive communication overhead required for encrypted computation, which affects both non-linear operations (e.g., ReLU approximation via comparison protocols) and linear operations (e.g., matrix-vector multiplication via share-wise computation and reconstruction)~\cite{zeng2025towards}. Specifically, each layer evaluation requires multiple rounds of interaction between parties to exchange intermediate shares, verify correctness, and reconstruct partial results~\cite{juvekar2018gazelle,wagh2019securenn,zeng2025towards}. As model depth and width increase, the cumulative communication volume grows substantially, creating significant deployment challenges in bandwidth-constrained mobile-cloud settings. Furthermore, SMPC approaches typically assume different system models depending on party roles: two-party protocols may assume computation between client and server~\cite{juvekar2018gazelle}, while three-party protocols often assume computation across multiple non-colluding servers~\cite{wagh2019securenn}. These requirements introduce additional challenges, including trust distribution, network coordination, and synchronization overhead, which are difficult to satisfy in real-world deployments
with heterogeneous devices and dynamic network conditions.

To address the above challenges, fully homomorphic encryption (FHE) has emerged as a strong alternative that eliminates multi-party coordination by enabling the server to compute directly on ciphertexts without accessing plaintext inputs~\cite{manh2025iot,acar2018survey}. Recent studies in FHE-based secure DL inference have focused on making encrypted image classification more efficient by optimizing FHE-encrypted linear and encrypted non-linear functions in neural networks (i.e., reducing the costly homomorphic operations such as ciphertext rotation and key switching)~\cite{zeng2025towards}, especially for CNN-based image classification models. However, despite these improvements, existing end-to-end FHE-based inference pipelines remain computationally intensive and memory-demanding, especially for deep models~\cite{zeng2025towards}. The need to evaluate all layers of a neural network in the encrypted domain results in high latency, large ciphertext sizes, and significant computational overhead. These limitations make current FHE-based approaches difficult to deploy in low-latency mobile-cloud environments.

\subsection{Related Works}

Representative HE/FHE systems for image classification have primarily focused on improving efficiency through packing strategies and operator-level scheduling. CryptoNets~\cite{gilad2016cryptonets} was among the first to demonstrate that neural network inference can be performed entirely on encrypted inputs. Using a leveled HE scheme with batched slot packing, CryptoNets evaluated a simple CNN on MNIST and achieved 99\% accuracy with a throughput of approximately 59,000 predictions per hour. Although the model was shallow and the per-image latency was high (around 250\,s on a single CPU core), it established the foundational paradigm of encoding neural network weights as plaintexts and evaluating layers through ciphertext-plaintext operations (e.g., multiplications and additions).

To reduce the computational overhead of deep encrypted CNNs, the authors in~\cite{lee2022low} proposed multiplexed parallel convolutions under the CKKS scheme. The proposed approach packed multiple channel outputs compactly into a single ciphertext to minimize the number of costly homomorphic operations (i.e., ciphertext rotation and bootstrapping). In addition, the authors integrated an imaginary-removing bootstrapping variant to prevent error divergence in approximated ReLU evaluation. On CIFAR-10 with ResNet-20, the study reported 2,271\,s latency for one encrypted image using a single CPU thread and 79\,s amortized runtime per image when classifying 50 images with 50 threads, corresponding to 4.67\,$\times$ lower inference latency and 134\,$\times$ less amortized runtime than the prior CKKS-based ResNet-20 implementation. Despite these improvements, the reported latency remains far above low-latency mobile-cloud requirements, and the framework still requires evaluating the full CNN depth under encryption.

At a larger model scale, NeuJeans~\cite{neujeans2024ccs} targeted the high cost of encrypted CNN inference by introducing Coefficients-in-Slot (CinS) encoding, which enables multiple convolutions within a single HE multiplication and reduces expensive slot permutations. The authors further showed that CinS aligns with the initial stages of the discrete Fourier transform (DFT), which is also used in CKKS bootstrapping~\cite{cheon2018bootstrapping}, thereby avoiding redundant CinS/slot conversion overhead in Conv2D execution flows. With these co-designed optimizations, NeuJeans reported up to 5.68\,$\times$ speedup for Conv2D-activation sequences and demonstrated ImageNet-scale encrypted CNN inference in a few seconds~\cite{neujeans2024ccs}. It also reported top-1/top-5 encrypted accuracies on 3,000 validation images (e.g., 74.1\%/90.5\% for ResNet-50). However, the reported end-to-end server inference delays are in the second-level range, i.e., 14.11\,s (ResNet-18 poly approximation), 56.08\,s (ResNet-50), and 40.76\,s (MobileNetV2), which remains too high for low-latency mobile-cloud applications.~\cite{neujeans2024ccs}.

Another recent direction focuses on channel-wise ciphertext organization for encrypted CNN inference. In this line,~\cite{cheon2025batch} proposed channel-by-channel packing and channel-by-channel convolution for CIFAR-10 classification with ResNet-20 and ResNet-110, and reported approximately 5.04\,$\times$ and 5.20\,$\times$ speedups with negligible accuracy deviation from plaintext. These gains are meaningful, but the deployment overhead remains high because the approach still executes a full deep CNN pipeline in the encrypted domain. In end-to-end HE-based inference, each convolutional layer requires numerous ciphertext rotations and key-switching operations to align packed slots and aggregate intermediate results
across channels and spatial dimensions. As model depth increases, the number of required rotation indices also grows, forcing the server to preload a large set of evaluation keys, particularly rotation keys, which substantially increases memory and storage overhead. Despite the packing optimizations, the study still reported an amortized runtime of 15.76\,s per image (batch size 512), around 370\,GB RAM usage, and 6.91\,GB of rotation-key storage.

{\color{blue}
While effective in principle, existing end-to-end HE-based DL approaches suffer from high latency and heavy computational and memory overhead. Compared with these approaches, split neural networks offer a promising way to reduce inference latency by executing different parts of a model at different entities. Specifically, Vepakomma~\emph{et al.}~\cite{vepakomma2018split} introduced SplitNN, in which a neural network is partitioned between participating entities so that raw data need not be directly shared. However, conventional SplitNN relies on data and model partitioning rather than cryptographic protection, i.e., intermediate representations are transmitted in plaintext, may retain sensitive input information, and can be exploited to reconstruct the original input~\cite{pasquini2021tiger,privascissors2023}. Alternative privacy-enhanced split-computing approaches protect these representations through different mechanisms. Pham~\emph{et al.}~\cite{pham2025dpsplit} integrated differential privacy into split learning by perturbing intermediate representations at the cut layer and investigated the resulting privacy-accuracy and communication trade-offs on image-classification tasks. This approach provides a statistical privacy guarantee without cryptographic computation. However, stronger perturbation can reduce predictive utility. Moreover, its primary focus is collaborative training rather than low-latency online inference. Khan~\emph{et al.}~\cite{khan2025splithappens} combined U-shaped split learning with function secret sharing, additive secret sharing, and Beaver triples to protect intermediate computations and labels. This design provides cryptographic confidentiality without perturbing the data. However, it primarily targets secure training and requires interactive computation between two non-colluding servers. These requirements complicate deployment in a latency-sensitive setting with a single cloud provider.

HE-enhanced split inference avoids plaintext-representation leakage and noise-induced utility loss, but introduces different system costs. For example, Pereteanu~\emph{et al.}~\cite{pereteanu2022splithe} partitioned a network into encrypted server-side front and back segments surrounding a larger plaintext middle segment executed by the client. This design substantially reduces encrypted computation compared with pure HE inference and protects the server-held model segments. However, each request must traverse the client-server boundary multiple times for decryption, plaintext middle-model evaluation, re-encryption, and further encrypted evaluation. Consequently, the design retains multi-second inference latency and increases the number of online communication rounds. Another example is the AI-as-a-Service architecture proposed by Lam~\emph{et al.}~\cite{lam2024efficient}. In this architecture, the user runs a public pretrained feature extractor locally, encrypts the resulting feature vector, and sends it to a cloud-hosted private network. The private network typically contains one or two fully connected layers evaluated under fully homomorphic encryption. The design encrypts each input feature separately and uses a look-up-table mechanism for nonlinear operations. To support these operations, the client must generate additional ciphertexts, evaluation keys, and key-switching material. These requirements increase client-side encryption, setup, and communication overhead. The cloud-side computation also increases as the feature dimension or private network grows. Collectively, these costs limit the deployability of existing privacy-preserving inference approaches in low-latency mobile-cloud environments and motivate the need for a more efficient system design.
}

\subsection{Proposed Approach and Main Contributions}

To overcome the above limitations in latency-sensitive mobile-cloud deployment, we propose PRICE, a \textbf{PR}ivacy-preserving \textbf{I}mage \textbf{C}lassification framework for mobile \textbf{E}dge-cloud computing. The design partitions inference across client, trusted edge, and untrusted cloud to minimize encrypted computation while preserving confidentiality. In online operation, the trusted edge performs deep feature extraction in plaintext, encrypts the resulting embedding with CKKS, and sends only ciphertext features to the cloud. The cloud executes ciphertext-only final-layer classification and returns encrypted classification results for decryption at the edge. By avoiding end-to-end encrypted deep-network execution at the cloud, this design reduces computational complexity, evaluation-key burden, and end-to-end latency. \textcolor{blue}{Unlike conventional SplitNN~\cite{vepakomma2018split}, our framework does not transmit plaintext intermediate representations between the edge and cloud. It also avoids the intermediate decryption, plaintext evaluation, and re-encryption steps required by Split HE~\cite{pereteanu2022splithe}. Therefore, our approach uses only one online edge-cloud request-response exchange. Our proposed framework has an architecture similar to that of Lam~\emph{et al.}~\cite{lam2024efficient}: both designs extract features before sending encrypted features to the cloud. However, their cloud-side computations are different. Our framework restricts cloud computation to encrypted final linear-layer evaluation. This restricted design enables RowHyDiag to optimize the encrypted matrix-vector computation. In contrast, Lam~\emph{et al.} support a more general private network that may include nonlinear layers evaluated using encrypted look-up tables.}

\textcolor{blue}{Under our intended managed multi-client deployment, the edge node operates within an organizational boundary, such as an enterprise or hospital gateway or a carrier-operated multi-access edge computing site. The edge is under local administrative control, whereas the third-party cloud remains outside the data owner's trust domain. Although modern mobile devices may have sufficient computational capability for local feature extraction, the edge in PRICE serves a broader role as a shared trust and service boundary rather than merely as a computational accelerator. It handles CKKS encryption and decryption, keeps the feature extractor and HE context persistently available, and amortizes model loading and cryptographic setup across multiple clients. This design avoids distributing secret keys and model copies to individual client devices. It also provides stable resources as the model size and number of concurrent requests grow. Accordingly, the cloud is untrusted and cannot access raw data, plaintext embeddings, or the secret key. Compared with fully encrypted neural-network inference, PRICE performs the computationally intensive feature extraction at the trusted edge and leaves only encrypted final linear-layer evaluation to the cloud.} To further improve cloud efficiency, we introduce RowHyDiag, a row-packed hybrid diagonal encoding with hoisted-rotation-aware encrypted linear-layer evaluation, which increases slot utilization and reduces redundant homomorphic operations. \textcolor{blue}{Unlike NeuJeans~\cite{neujeans2024ccs}, which improves end-to-end encrypted neural-network inference by co-optimizing convolution and FHE bootstrapping through the CinS encoding, our proposed approach RowHyDiag targets efficient encrypted evaluation of the final linear classifier executed by the cloud after the trusted edge has completed plaintext feature extraction. Its algorithmic novelty lies in jointly designing the row-packed ciphertext layout, diagonal encoding, hoisted rotation schedule, and structured reduction so that multiple output rows are evaluated concurrently within the same CKKS slot space. This joint design reduces the number of ciphertext rotations by up to $85.3\times$ across the evaluated 128-bit-security configurations.} Together, the trusted-edge partitioning and RowHyDiag reduce the encrypted workload to a shallow final-layer computation, lowering ciphertext rotations, evaluation-key usage, cloud setup overhead, and end-to-end inference latency while maintaining predictive accuracy across multiple image-classification workloads.

\begin{figure*}[t]
\centering
\includegraphics[width=0.95\textwidth]{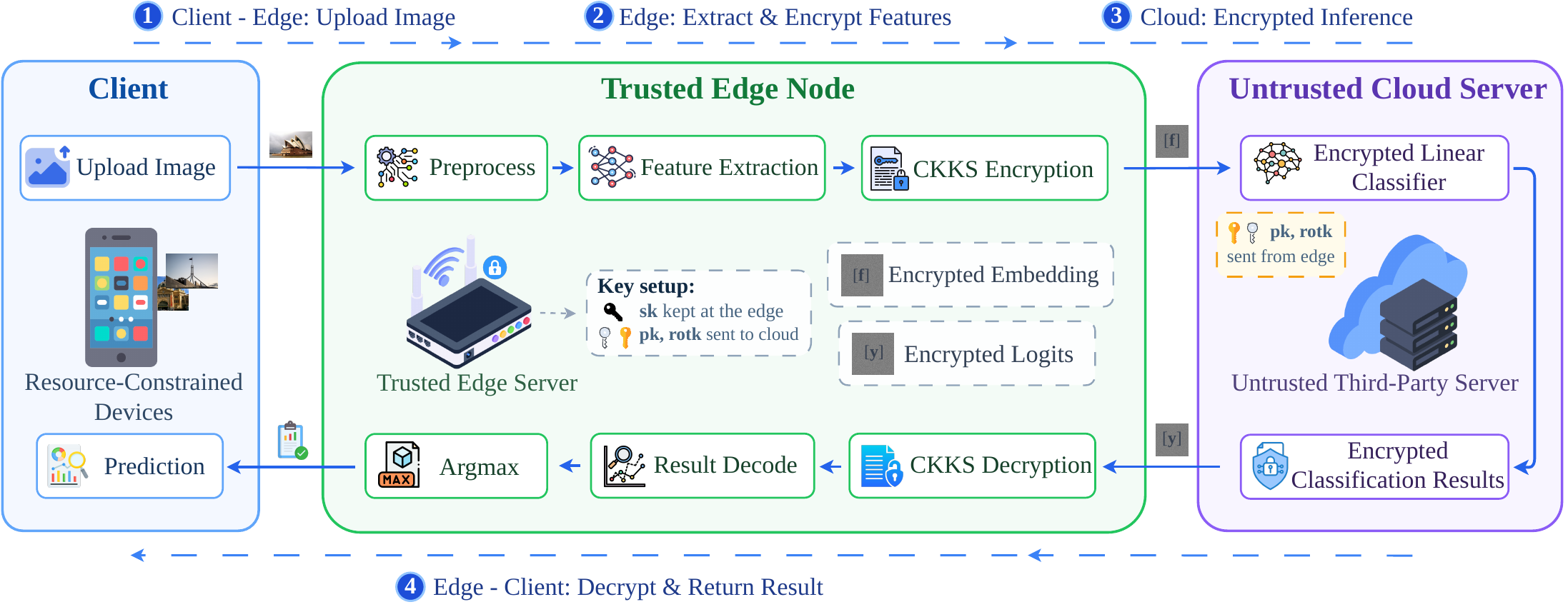}
\caption{System architecture showing three-tier workflow: (1) client sends image to edge server, (2) edge server extracts and encrypts features, (3) cloud server performs encrypted inference, (4) edge server decrypts and returns result to client.}
\label{fig:framework}
\end{figure*}

Specifically, the main contributions of this work include:
\begin{itemize}
        \item We propose a three-tier edge-cloud architecture that separates plaintext feature extraction and encrypted classification. By executing deep feature extraction at a trusted edge and restricting homomorphic computation to the final classifier at the untrusted cloud, our design \textcolor{blue}{avoids evaluating multiple neural-network layers under encryption and substantially} reduces encrypted computational complexity and end-to-end inference delay \textcolor{blue}{under an explicitly stated trusted-edge threat model}.

        \item We develop RowHyDiag for efficient encrypted linear-layer computation in image classification. \textcolor{blue}{Across the evaluated 128-bit-security configurations, RowHyDiag achieves 12.2-31.8$\times$ faster cloud inference, 32.0-85.3$\times$ fewer rotations, and 31.6-83.0$\times$ lower one-time cryptographic setup payload than the conventional Halevi-Shoup diagonal method. It also outperforms the PEGASUS linear-transformation baseline, providing 1.1-4.5$\times$ faster cloud inference and 1.2-3.4$\times$ lower setup payload. These improvements are validated through theoretical analysis and empirical evaluation.}

        \item We conduct extensive simulations and real-device experiments on benchmark datasets (MNIST, CIFAR-10, CIFAR-100, \textcolor{blue}{ImageNet-1K, and Australian Landmarks}). Results show strong predictive fidelity and deployment-ready performance: \textcolor{blue}{encrypted accuracy consistently matches plaintext accuracy across all five datasets under different HE parameter configurations, while the maximum relative logit error remains between $9.67\times10^{-9}$ and $3.22\times10^{-4}$ across all evaluated settings.}

        \item We additionally build a new application-oriented Australian Landmarks dataset to better evaluate the performance of the proposed approach on real image classification tasks. To the best of our knowledge, this is the first curated Australian-focused landmark dataset used in prior privacy-preserving inference studies, providing an essential real-image benchmark for evaluating system performance under realistic deployment conditions.
\end{itemize}

The remainder of this paper is organized as follows. Section~\ref{sec:framework} describes the proposed three-tier framework with main components. Section~\ref{sec:technique} details the privacy-preserving image classification technique and Section~\ref{sec:setup} presents the datasets, evaluation metrics, experimental configuration, and baseline methods for evaluation. Section~\ref{sec:results} reports simulation and low-latency edge-cloud results and Section~\ref{sec:conclusion} concludes the paper.

\section{The Proposed Framework}
\label{sec:framework}

Fig.~\ref{fig:framework} presents the proposed three-tier architecture for privacy-preserving image classification. The system comprises: (i) resource-constrained client devices (e.g., smartphones and tablets), (ii) trusted edge servers deployed within an organizational boundary, and (iii) third-party cloud servers treated as untrusted. The core design principle is to ensure that raw user images never leave the trusted environment, while delegating only encrypted computation to the cloud. In this workflow, client devices transmit images to the trusted edge, which performs preprocessing and deep feature extraction in plaintext, then encrypts the resulting embeddings using the CKKS scheme~\cite{cheon2017ckks}. The untrusted cloud only executes encrypted inference over ciphertexts, and encrypted outputs are returned to the edge, where the returned ciphertext is decrypted, and the final classification result is then sent back to the user. This three-tier architecture preserves data confidentiality while enabling scalable cloud-assisted computation.

Before online inference, the system executes a one-time key setup phase. The trusted edge server generates CKKS keys, including a public key $\mathbf{pk}$ for encryption, a secret key $\mathbf{sk}$ for decryption, and evaluation keys (e.g., rotation keys $\mathbf{rotk}$) for homomorphic operations. The public and evaluation keys ($\mathbf{pk}$, $\mathbf{rotk}$) are provisioned to the cloud server, while the secret key $\mathbf{sk}$ remains exclusively at the edge.

\subsection{Client Device}
In the intended managed multi-client deployment, the client is responsible for data acquisition and transmits the selected image to a nearby trusted edge through a secure HTTPS channel~\cite{rfc8446}.

\subsection{Trusted Edge Node}
The trusted edge node handles two main processes: (i) feature extraction and encryption before cloud offloading, and (ii) decryption and final prediction after cloud inference. During deployment, the full model is first fine-tuned offline, then partitioned such that the feature extractor is retained at the edge while the final linear classifier is deployed to the cloud.

\subsubsection{Extract and Encrypt Features}
Upon receiving an image from the client, the edge performs preprocessing and runs the deployed feature extractor to obtain a fixed-dimensional embedding. The edge device then applies $\ell_2$ normalization to stabilize encrypted computation~\cite{schroff2015facenet} and encrypts the normalized feature vector using CKKS. Only ciphertext embeddings are transmitted to the cloud, ensuring that raw images and plaintext features never leave the trusted environment. The process is explained in detail in Section~\ref{subsec:edge_preparation_encryption}.

\subsubsection{Decrypt and Return Results}
After receiving encrypted classification results from the cloud, the edge device decrypts them with the secret key, decodes the recovered plaintext class scores, applies argmax, and returns the final prediction to the client. All decryption and post-processing operations are confined to the trusted edge, preserving data confidentiality. The process is explained in detail in Section~\ref{subsec:edge_decrypt_recover}.

\subsection{Untrusted Cloud Server}
The untrusted cloud server stores and executes only the encrypted final-layer classifier, together with CKKS public/evaluation keys. It performs encrypted linear-layer inference (matrix-vector multiplication with bias addition) over received encrypted embeddings, then returns encrypted class-score ciphertexts to the edge. Since the cloud never has access to the secret key or plaintext data, it can scale inference while preserving data confidentiality. The process is further detailed in Section~\ref{subsec:cloud_ciphertext_inference}.

{\color{blue}
\subsection{Threat Model Scope and Limitations}
Our framework considers scenarios in which the edge node is trusted, i.e., it can access raw data, plaintext feature embeddings, model-backbone parameters, and the CKKS secret key. Following the ETSI security architecture standard, the edge operates inside an organizational security boundary, such as an enterprise or hospital gateway, an on-premise industrial appliance, or a carrier-operated MEC site governed by local administrative control and platform-attestation policies~\cite{etsimec041_2024,etsimec003_2024}.

In scenarios where the edge node cannot be fully trusted, plaintext feature extraction and cryptographic key management should be performed within a trusted execution environment (TEE), such as Intel SGX~\cite{cowan2017sgx} or ARM TrustZone~\cite{sabt2015trustzone}. The client could first verify the TEE through remote attestation and establish a secure channel terminating inside the TEE. This channel thus can protect the data using conventional encryption method at the transport layer. Furthermore, the edge OS will only forward encrypted traffic, while the data will be decrypted and processed inside hardware-protected memory (i.e., TEE). Feature extraction and cryptographic key management would also be performed inside the TEE. Therefore, software outside the TEE, including a compromised host OS or hypervisor, would not have access to raw data, plaintext embeddings, or cryptographic keys. Incorporating a TEE would inevitably introduce additional overhead associated with enclave memory constraints, context switching, model loading, and remote attestation. Nevertheless, TEE-based protection complements rather than alters the cryptographic design of RowHyDiag. A comprehensive integration and evaluation of TEEs is therefore left for future deployment-oriented work.
}

\section{Proposed Privacy-Preserving Image Classification Technique}
\label{sec:technique}

In this section, we first provide fundamental background on HE and its core operations. We then describe in detail the proposed methods and algorithms introduced in Section~\ref{sec:framework}.

\subsection{Homomorphic Encryption Background}
\label{subsec:homomorphic_encryption}

Homomorphic encryption is a cryptographic approach that allows mathematical operations to be performed directly on encrypted data without decryption. To work with real numbers in neural network layers, we use the CKKS scheme~\cite{cheon2017ckks}, which supports approximate arithmetic on floating-point values. Additionally, CKKS enables Single Instruction, Multiple Data (SIMD)-style processing, which allows multiple data elements to be encoded within a single ciphertext~\cite{chen2018simdbatching}, thereby enabling efficient parallel computation for linear layers in neural networks. Additionally, the CKKS scheme is based on the Ring Learning With Errors (RLWE) problem~\cite{lyubashevsky2010rlwe}, providing strong privacy against decryption attempts. The scheme requires ciphertexts to be represented as one-dimensional vectors over a ring of dimension $N$, with the number of available slots equal to $N/2$. The ring dimension $N$ determines both the security level and the supported multiplicative depth. Importantly, raw data must be encoded into plaintexts before encryption so that homomorphic operations can be applied. The main homomorphic operations used in our encrypted inference pipeline are:

\begin{itemize}
    \item \textbf{Key Generation}: Generates $\mathbf{sk}$, $\mathbf{pk}$, and rotation keys $\mathbf{rotk}$ for computation.

    \item \textbf{Encryption}: Given a plaintext vector $\mathbf{c} \in \mathbb{R}^{d}$ and $\mathbf{pk}$, encrypt $\mathbf{c}$ into a ciphertext $\overline{c}$ where $\overline{c} = \operatorname{Encrypt}(\mathbf{pk}, \mathbf{c})$.

    \item \textbf{Decryption}: Given a ciphertext $\overline{c}$ and secret key $\mathbf{sk}$, decrypt $\overline{c}$ into its plaintext form $\mathbf{c}$ where $\mathbf{c} = \operatorname{Decrypt}(\mathbf{sk}, \overline{c})$.

    \item \textbf{Addition}: Given two ciphertexts $\overline{c}_1$ and $\overline{c}_2$, perform element-wise encrypted addition $\operatorname{EvalAdd}(\overline{c}_1, \overline{c}_2) = \overline{c}_{\text{add}}$, where $\operatorname{Decrypt}(\mathbf{sk}, \overline{c}_{\text{add}}) \approx \mathbf{c}_1 + \mathbf{c}_2$.

    \item \textbf{Ciphertext-Plaintext Addition}: Given a ciphertext $\overline{c}_1$ and a plaintext $\mathbf{c}_2$, perform element-wise encrypted addition $\operatorname{EvalAdd}(\overline{c}_1, \mathbf{c}_2) = \overline{c}_{\text{add}}$, where $\operatorname{Decrypt}(\mathbf{sk}, \overline{c}_{\text{add}}) \approx \mathbf{c}_1 + \mathbf{c}_2$.

    \item \textbf{Multiplication}: Given two ciphertexts $\overline{c}_1$ and $\overline{c}_2$, perform element-wise encrypted multiplication $\operatorname{EvalMult}(\overline{c}_1, \overline{c}_2) = \overline{c}_{\text{mult}}$, where $\operatorname{Decrypt}(\mathbf{sk}, \overline{c}_{\text{mult}}) \approx \mathbf{c}_1 \times \mathbf{c}_2$.

    \item \textbf{Ciphertext-Plaintext Multiplication}: Given a ciphertext $\overline{c}_1$ and a plaintext $\mathbf{c}_2$, perform element-wise encrypted multiplication $\operatorname{EvalMult}(\overline{c}_1, \mathbf{c}_2) = \overline{c}_{\text{mult}}$, where $\operatorname{Decrypt}(\mathbf{sk}, \overline{c}_{\text{mult}}) \approx \mathbf{c}_1 \times \mathbf{c}_2$.

    \item \textbf{Rotation}: Given a ciphertext $\overline{c}$ encrypting a vector $(\mathbf{c}_0, \ldots, \mathbf{c}_{\frac{N}{2}-1})$, perform rotation by $i$ positions: $\operatorname{EvalRotate}(\overline{c}, i) = \overline{c}_{\text{rot}}$, where $\operatorname{Decrypt}(\mathbf{sk}, \overline{c}_{\text{rot}}) \approx (\mathbf{c}_i, \ldots, \mathbf{c}_{\frac{N}{2}-1}, \mathbf{c}_0, \ldots, \mathbf{c}_{i-1})$~\cite{halevi2014algorithms}.

    \item \textbf{Hoisted Rotation}: For efficient multiple rotations of the same ciphertext, hoisted key switching~\cite{chen2019hoisted} precomputes $\text{hoisted} = \operatorname{EvalFastRotationPrecompute}(\overline{c})$ once, then performs $\overline{c}_{\text{rot}} = \operatorname{EvalFastRotation}(\overline{c}, i, \text{hoisted})$ for each rotation index $i$.

    \item \textbf{Sum Across Slots}: Given a ciphertext $\overline{c}$ with batch size $B$, perform $\operatorname{EvalSum}(\overline{c}, B) = \overline{c}_{\text{sum}}$ using a binary-tree reduction pattern: in each round $r=0,1,\ldots,\lceil \log_2 B \rceil - 1$, the ciphertext is rotated by $2^r$ positions and added to itself, progressively halving the number of unaccumulated elements. After $\lceil \log_2 B \rceil$ rounds, the sum of all $B$ slot values is replicated across all positions in the resulting ciphertext.
\end{itemize}

Building on these CKKS primitives, the following subsections describe the end-to-end encrypted inference workflow, including feature extraction and encryption at the edge, encrypted inference in the cloud, and decryption
with result recovery at the edge (Fig.~\ref{fig:framework} steps 2-4).

\subsection{Feature Extraction and Encryption at the Trusted Edge}
\label{subsec:edge_preparation_encryption}

The trusted edge receives an image from the client, performs preprocessing and feature extraction in plaintext, and encrypts only the resulting feature embedding before offloading it to the cloud. This subsection formalizes the query encoding process and encryption procedure.

Given an input image $\mathbf{x} \in \mathbb{R}^{H \times W \times C}$ where $H$, $W$, and $C$ denote height, width, and number of channels, the trusted edge first applies standard preprocessing operations (e.g., resizing, normalization) to obtain $\mathbf{x}' = \mathcal{P}(\mathbf{x})$, where $\mathcal{P}: \mathbb{R}^{H \times W \times C} \to \mathbb{R}^{H' \times W' \times C'}$ represents the preprocessing transformation. The preprocessed image is then fed into a pre-trained feature extractor $\Phi: \mathbb{R}^{H' \times W' \times C'} \to \mathbb{R}^{m}$ to extract a compact feature embedding:
\begin{equation}
\mathbf{f} = \Phi(\mathbf{x}') = \Phi(\mathcal{P}(\mathbf{x})) \in \mathbb{R}^{m},
\label{eq:feature_embedding_pipeline}
\end{equation}
where $m$ is the feature dimension (e.g., $m=512$ for ResNet-34, $m=768$ for ViT-B/16). This embedding $\mathbf{f} = [f_0, f_1, \ldots, f_{m-1}]^\top$ serves as the plaintext input for subsequent encrypted classification.

For efficient encrypted computation using CKKS with ring dimension $N$, each ciphertext provides $S = N/2$ available slots. To improve slot utilization and support efficient encrypted computation (especially tree-style reduction during encrypted inference), we design a slot-replication encoding strategy. First, we pad the feature dimension to the nearest power of two:
\begin{equation}
m_{\text{padded}} = 2^{\lceil \log_2 m \rceil},
\label{eq:query_m_padded}
\end{equation}
which ensures compatibility with binary-tree reduction patterns. The padded feature vector is defined as:
\begin{equation}
\mathbf{f}_{\text{padded}} = [f_0, f_1, \ldots, f_{m-1}, \underbrace{0, \ldots, 0}_{m_{\text{padded}} - m}] \in \mathbb{R}^{m_{\text{padded}}}.
\label{eq:query_f_padded}
\end{equation}

Given that $m_{\text{padded}} \leq S$ (a condition satisfied by typical configurations where $m \ll S$), we compute the number of feature replicas that fit into a single ciphertext. Since each replica occupies $m_{\text{padded}}$ consecutive slots, the total number of replicas is obtained by dividing the available slot count $S$ by the replica size $m_{\text{padded}}$ and taking the floor to discard any incomplete trailing block:
\begin{equation}
n_{\text{pack}} = \left\lfloor \frac{S}{m_{\text{padded}}} \right\rfloor.
\label{eq:query_n_pack}
\end{equation}
We then replicate $\mathbf{f}_{\text{padded}}$ across $n_{\text{pack}}$ consecutive slot blocks to form the replicated feature vector:
\begin{equation}
\mathbf{f}_{\text{rep}} = [\underbrace{\mathbf{f}_{\text{padded}}, \mathbf{f}_{\text{padded}}, \ldots, \mathbf{f}_{\text{padded}}}_{n_{\text{pack}} \text{ copies}}] \in \mathbb{R}^{n_{\text{pack}} \cdot m_{\text{padded}}}.
\label{eq:query_frep}
\end{equation}
This replication strategy enables efficient homomorphic matrix-vector multiplication within our proposed encrypted inference algorithm, which is detailed in Section~\ref{subsec:cloud_ciphertext_inference}. Finally, the edge encrypts the replicated feature vector using the CKKS public key:
\begin{equation}
\overline{\mathbf{f}}_{\text{rep}} = \operatorname{Encrypt}(\mathbf{pk}, \mathbf{f}_{\text{rep}}),
\label{eq:query_encrypted}
\end{equation}
resulting in an encrypted query ciphertext $\overline{\mathbf{f}}_{\text{rep}}$ that is transmitted to the untrusted cloud for encrypted inference. For illustration, Fig.~\ref{fig:query_encoding} in \appref{app:encoding_illustrations} describes this query encoding workflow, including power-of-two padding and slot replication.

\subsection{Encrypted Inference in the Untrusted Cloud}

In standard plaintext inference, the linear classification layer computes class logits via a straightforward matrix-vector multiplication followed by bias addition. Given a feature vector $\mathbf{f} \in \mathbb{R}^{m}$, weight matrix $\mathbf{W} \in \mathbb{R}^{n \times m}$, and bias vector $\mathbf{b} \in \mathbb{R}^{n}$, where $n$ denotes the number of classes and $m$ is the feature dimension, the linear layer computes:
\begin{equation}
\boldsymbol{\ell} = \mathbf{W} \mathbf{f} + \mathbf{b} \in \mathbb{R}^{n},
\label{eq:plaintext_linear}
\end{equation}
where $\boldsymbol{\ell} = [\ell_0, \ell_1, \ldots, \ell_{n-1}]^\top$ represents the output logits. The predicted class is then determined by $\hat{y} = \arg\max_{i} \ell_i$. In conventional plaintext computation, this matrix-vector product is readily computed via standard linear algebra operations.

\label{subsec:cloud_ciphertext_inference}

\begin{algorithm}[t]
\caption{RowHyDiag Encoding: Pack rows and encode diagonals}
\label{alg:rowpack}
\begin{algorithmic}[1]
\STATE \textbf{Input:} Weight matrix $\mathbf{W} \in \mathbb{R}^{n \times m}$, batch size $S$
\STATE \textbf{Output:} Packed matrix $\mathbf{W}_{\text{packed}} \in \mathbb{R}^{n_{\text{padded}} \times S}$, diagonal plaintexts $\{\text{diag}_d\}_{d=0}^{n_{\text{padded}}-1}$
\STATE Compute $m_{\text{padded}} \leftarrow 2^{\lceil \log_2 m \rceil}$
\STATE Compute $n_{\text{pack}} \leftarrow \lfloor S/m_{\text{padded}} \rfloor$
\STATE Compute $n_{\text{packed}} \leftarrow \lceil n / n_{\text{pack}} \rceil$
\STATE Compute $n_{\text{padded}} \leftarrow 2^{\lceil \log_2 n_{\text{packed}} \rceil}$
\FOR{$i_{\text{packed}} = 0$ \TO $n_{\text{padded}} - 1$}
	\FOR{$j = 0$ \TO $S - 1$}
		\STATE $i_{\text{orig}} \leftarrow i_{\text{packed}} n_{\text{pack}} + \lfloor j/m_{\text{padded}} \rfloor$
		\STATE $j_{\text{orig}} \leftarrow j \bmod m_{\text{padded}}$
		\IF{$i_{\text{orig}} < n$ \AND $j_{\text{orig}} < m$}
			\STATE $\mathbf{W}_{\text{packed}}[i_{\text{packed}}, j] \leftarrow \mathbf{W}[i_{\text{orig}}, j_{\text{orig}}]$
		\ELSE
			\STATE $\mathbf{W}_{\text{packed}}[i_{\text{packed}}, j] \leftarrow 0$
		\ENDIF
	\ENDFOR
\ENDFOR
\FOR{$d = 0$ \TO $n_{\text{padded}} - 1$}
	\FOR{$j = 0$ \TO $S - 1$}
		\STATE $r \leftarrow j \bmod n_{\text{padded}}$
		\STATE $\text{blk} \leftarrow \lfloor j / m_{\text{padded}} \rfloor$
		\STATE $p \leftarrow j \bmod m_{\text{padded}}$
		\STATE $c \leftarrow \text{blk}\, m_{\text{padded}} + ((d + p) \bmod m_{\text{padded}})$
		\STATE $\text{diag}_d[j] \leftarrow \mathbf{W}_{\text{packed}}[r, c]$
	\ENDFOR
\ENDFOR
\STATE \textbf{return} $\mathbf{W}_{\text{packed}}, \{\text{diag}_d\}_{d=0}^{n_{\text{padded}}-1}$
\end{algorithmic}
\end{algorithm}

However, performing the linear forward pass in~\eqref{eq:plaintext_linear} homomorphically over CKKS ciphertexts introduces fundamental structural constraints. As described in Section~\ref{subsec:homomorphic_encryption}, CKKS ciphertexts require data to be encoded in one-dimensional vectors packed into polynomial slots, and the supported homomorphic operations (ciphertext-plaintext multiplication, ciphertext addition, and slot rotation) are inherently element-wise (Hadamard product) or permutation-based rather than standard matrix operations. Consequently, directly applying $\mathbf{W}$ to an encrypted vector $\overline{\mathbf{f}}$ is not feasible under the CKKS evaluation model. To enable efficient encrypted computation in~\eqref{eq:plaintext_linear}, the weight matrix $\mathbf{W}$ and bias vector $\mathbf{b}$ must be transformed into plaintext representations that align with the one-dimensional ciphertext slot structure and leverage the element-wise multiplication and rotation primitives provided by CKKS.

To address this challenge, we propose RowHyDiag (row-packed hybrid diagonal encoding), which transforms $\mathbf{W}$ and $\mathbf{b}$ into structured diagonal plaintexts compatible with rotation-and-multiply-based encrypted matrix-vector multiplication. The untrusted cloud receives only the encrypted feature representation $\overline{\mathbf{f}}_{\text{rep}}$ from the trusted edge and fully evaluates the linear mapping in~\eqref{eq:plaintext_linear} in the encrypted domain using ciphertext-plaintext operations. This subsection details the RowHyDiag encoding for the weight matrix and bias vector, followed by the encrypted evaluation procedure that realizes the same mapping in~\eqref{eq:plaintext_linear} under encryption.

\subsubsection{RowHyDiag Matrix Encoding}
\label{subsubsec:rowpacked_hybrid_encoding}

Given classifier weights $\mathbf{W}\in\mathbb{R}^{n\times m}$ and slot count $S$, we compute
\begin{equation}
n_{\text{packed}}=\left\lceil\frac{n}{n_{\text{pack}}}\right\rceil,
\label{eq:n_packed}
\end{equation}
\begin{equation}
n_{\text{padded}}=2^{\lceil\log_2 n_{\text{packed}}\rceil}.
\label{eq:n_padded}
\end{equation}

Starting from the original matrix of size $n\times m$, we first align columns with $m_{\text{padded}}$, then pack $n_{\text{pack}}$ original rows into each packed row to obtain a matrix of size $n_{\text{packed}}\times S$. We next zero-pad rows until reaching $n_{\text{padded}}$ (which equals $n_{\text{packed}}$ when $n_{\text{packed}}$ is already a power of two), making the packed matrix compatible with power-of-two scheduling in subsequent encrypted operations. This construction improves slot utilization while preserving the original linear mapping. For illustration, Fig.~\ref{fig:hybrid_encoding_main} in \appref{app:encoding_illustrations} summarizes the square $4\times 4$ workflow, whereas Fig.~\ref{fig:nonsquare_encoding_appendix} in \appref{app:nonsquare_examples} provides a detailed non-square example.

Rows of $\mathbf{W}$ are packed blockwise into $\mathbf{W}_{\text{packed}}\in\mathbb{R}^{n_{\text{padded}}\times S}$, and diagonal plaintexts $\{\text{diag}_d\}_{d=0}^{n_{\text{padded}}-1}$ are extracted for rotation-and-multiply evaluation according to the slot-index mapping, as formalized in~\eqref{eq:diag_row}-\eqref{eq:diag_col}.

For $d\in\{0,\ldots,n_{\text{padded}}-1\}$ and slot $j\in\{0,\ldots,S-1\}$, each diagonal element is defined by
\begin{gather}
r = j \bmod n_{\text{padded}}, \label{eq:diag_row}\\
\text{blk} = \left\lfloor j / m_{\text{padded}} \right\rfloor, \label{eq:diag_block}\\
p = j \bmod m_{\text{padded}}, \label{eq:diag_pos}\\
c = \text{blk}\cdot m_{\text{padded}} + \left((d+p) \bmod m_{\text{padded}}\right), \label{eq:diag_col}
\end{gather}
where $r$ denotes the row index, $\text{blk}$ denotes the block identifier, $p$ denotes the in-block position, and $c$ denotes the column index. This block-local wrap-around preserves index alignment between packed features and packed weights.

This encoding ensures that when we rotate $\overline{\mathbf{f}}_{\text{rep}}$ by $d$ positions and multiply with $\text{diag}_d$, we obtain the correct element-wise products for encrypted linear-layer evaluation. The proposed RowHyDiag encoding is detailed in Algorithm~\ref{alg:rowpack}.

\subsubsection{Bias Encoding}
\label{subsubsec:bias_encoding}

\begin{algorithm}[t]
\caption{BiasEncode: Encode aligned bias plaintext}
\label{alg:bias_encode}
\begin{algorithmic}[1]
\STATE \textbf{Input:} Bias $\mathbf{b} \in \mathbb{R}^{n}$, $S$, $m_{\text{padded}}$, $n_{\text{pack}}$, $n_{\text{packed}}$
\STATE \textbf{Output:} Bias plaintext $\text{bias}_{\text{pt}}$
\FOR{$j = 0$ \TO $S - 1$}
	\STATE $\text{blk} \leftarrow \lfloor j / m_{\text{padded}} \rfloor$
	\STATE $p \leftarrow j \bmod m_{\text{padded}}$
	\IF{$p < n_{\text{packed}}$}
		\STATE $i_{\text{bias}} \leftarrow p \cdot n_{\text{pack}} + \text{blk}$
		\IF{$i_{\text{bias}} < n$}
			\STATE $\text{bias}_{\text{pt}}[j] \leftarrow \mathbf{b}[i_{\text{bias}}]$
		\ELSE
			\STATE $\text{bias}_{\text{pt}}[j] \leftarrow 0$
		\ENDIF
	\ELSE
		\STATE $\text{bias}_{\text{pt}}[j] \leftarrow 0$
	\ENDIF
\ENDFOR
\STATE \textbf{return} $\text{bias}_{\text{pt}}$
\end{algorithmic}
\end{algorithm}

\begin{figure*}[t]
\centering
\includegraphics[width=1\textwidth]{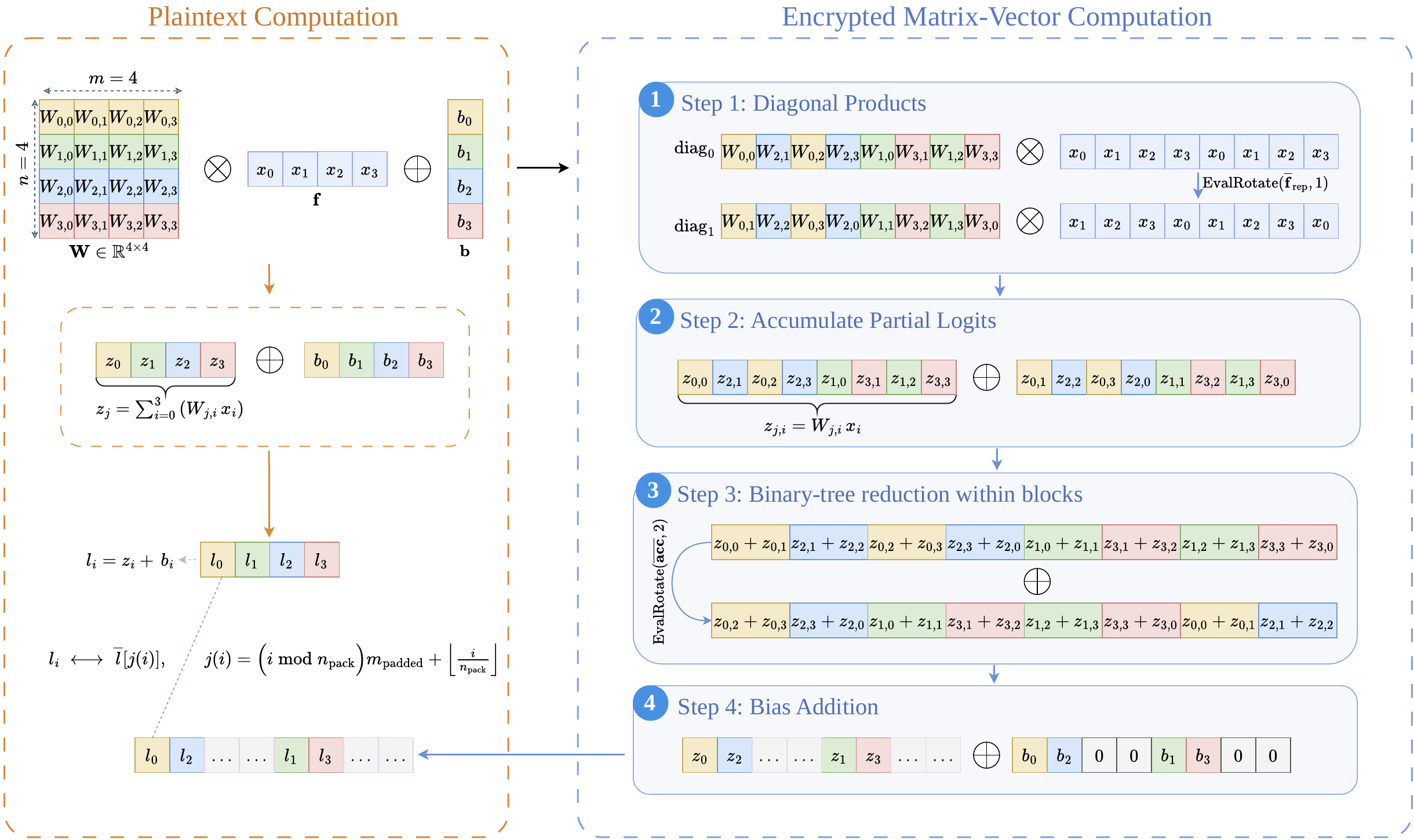}
\caption{Encrypted linear-layer evaluation workflow for Algorithm~\ref{alg:encrypted_inference}. Left (plaintext): the linear classifier computes $\boldsymbol{\ell}=\mathbf{W}\mathbf{f}+\mathbf{b}$. Right (Encrypted): step 1 forms diagonal products from rotated packed features and diagonal plaintexts; step 2 accumulates partial results; step 3 performs in-block binary-tree reduction via rotations and additions; step 4 adds the aligned bias plaintext to obtain encrypted results $\overline{\boldsymbol{\ell}}$. The final class results are decoded from packed slots using the slot-to-class index mapping.}
\label{fig:matrix_vector_computation}
\end{figure*}

To align post-multiplication accumulation with classifier biases, we encode $\mathbf{b}\in\mathbb{R}^{n}$ into a plaintext vector $\text{bias}_{\text{pt}}$ using the same block layout. For slot index $j$, let $\text{blk}=\lfloor j/m_{\text{padded}}\rfloor$ and $p=j\bmod m_{\text{padded}}$. The mapped class index is
\begin{equation}
i_{\text{bias}}=p\,n_{\text{pack}}+\text{blk},
\label{eq:bias_index_map}
\end{equation}
and $\text{bias}_{\text{pt}}[j]=\mathbf{b}[i_{\text{bias}}]$ if $i_{\text{bias}}<n$; otherwise zero. Equivalently, biases are distributed across $n_{\text{pack}}$ blocks of size $m_{\text{padded}}$, where block $k$ stores biases for rows $k, k+n_{\text{pack}}, k+2n_{\text{pack}},\ldots$ at the block prefix. This layout matches the organization of intermediate encrypted accumulation, enabling correct homomorphic bias addition without extra reshaping. Fig.~\ref{fig:bias_encoding_main} in \appref{app:encoding_illustrations} illustrates the square-case layout, while Fig.~\ref{fig:bias_encoding_appendix} in \appref{app:nonsquare_examples} shows a non-square example. The full procedure is provided in Algorithm~\ref{alg:bias_encode}.

\subsubsection{Encrypted Linear Layer Evaluation}
\label{subsubsec:encrypted_matvec}

Given the encoded weight diagonals $\{\text{diag}_d\}$, aligned bias plaintext $\text{bias}_{\text{pt}}$, and encrypted query $\overline{\mathbf{f}}_{\text{rep}}$ from the trusted edge, the untrusted cloud evaluates the linear layer in~\eqref{eq:plaintext_linear} homomorphically using rotations, ciphertext-plaintext multiplications, and additions.

We first perform hoisted key-switching precomputation for $\overline{\mathbf{f}}_{\text{rep}}$ to amortize the cost of multiple ciphertext rotations. For $d=0$, we multiply the unrotated ciphertext by $\text{diag}_0$ to initialize the accumulator; for each $d\in\{1,\ldots,n_{\text{padded}}-1\}$, we rotate the encrypted query by $d$, multiply by $\text{diag}_d$, and accumulate via homomorphic addition. This produces partial results distributed across $n_{\text{pack}}$ blocks. \textcolor{blue}{Because the packed-row coordinate has period $n_{\text{padded}}$,} we then apply in-block binary-tree reduction with strides \textcolor{blue}{$n_{\text{padded}},2n_{\text{padded}},4n_{\text{padded}},\ldots$}, and finally add the aligned bias plaintext to obtain encrypted results $\overline{\boldsymbol{\ell}}$.

Formally, let $\rho^d(\cdot)$ denote left rotation by $d$ slots, and $\odot$ denote slot-wise multiplication. The diagonal accumulation computes:
\begin{equation}
\overline{\mathbf{u}}_d=\rho^d(\overline{\mathbf{f}}_{\text{rep}})\odot\text{diag}_d,
\label{eq:diag_product}
\end{equation}
\begin{equation}
\overline{\mathbf{a}}=\sum_{d=0}^{n_{\text{padded}}-1}\overline{\mathbf{u}}_d.
\label{eq:diag_accumulator}
\end{equation}
At the slot level:
\begin{equation}
a[j]=\sum_{d=0}^{n_{\text{padded}}-1}\text{diag}_d[j]f_{\text{rep}}[(j+d)\bmod S].
\label{eq:diag_slotwise}
\end{equation}
Here, $f_{\text{rep}}[j]$ denotes the $j$-th slot of $\operatorname{Decrypt}(\mathbf{sk},\overline{\mathbf{f}}_{\text{rep}})$.
Under this notation, the summation is exactly the blockwise wrapped diagonal accumulation induced by~\eqref{eq:diag_row}-\eqref{eq:diag_col}.

We then apply in-block binary-tree reduction:
\begin{equation}
\overline{\mathbf{a}}^{(0)}=\overline{\mathbf{a}},\qquad
\textcolor{blue}{T=\left\lceil\log_2\left(\frac{m_{\text{padded}}}{n_{\text{padded}}}\right)\right\rceil,}
\label{eq:reduction_init}
\end{equation}
\begin{equation}
\textcolor{blue}{\overline{\mathbf{a}}^{(t+1)}=\overline{\mathbf{a}}^{(t)}+\rho^{2^t n_{\text{padded}}}\!\left(\overline{\mathbf{a}}^{(t)}\right),\quad t=0,\ldots,T-1.}
\label{eq:reduction_round}
\end{equation}
Finally,
\begin{equation}
\overline{\boldsymbol{\ell}}=\overline{\mathbf{a}}^{(T)}+\text{bias}_{\text{pt}}.
\label{eq:final_bias_add}
\end{equation}

For detailed illustration, Fig.~\ref{fig:matrix_vector_computation} visualizes the end-to-end encrypted evaluation flow corresponding to~\eqref{eq:diag_product}-\eqref{eq:final_bias_add}. Algorithm~\ref{alg:encrypted_inference} summarizes the full cloud-side encrypted evaluation procedure.

\subsection{Decryption and Result Recovery at the Trusted Edge}
\label{subsec:edge_decrypt_recover}

After cloud evaluation, encrypted results are returned to the trusted edge for decryption and class-score extraction. The trusted edge first decrypts the returned ciphertext using the secret key:
\begin{equation}
\mathbf{a}_{\text{dec}} = \operatorname{Decrypt}(\mathbf{sk},\overline{\boldsymbol{\ell}}) \in \mathbb{R}^{S},
\label{eq:decryption}
\end{equation}
where $\mathbf{a}_{\text{dec}}$ is the decrypted plaintext vector containing the packed logit values across $S$ slots.

\begin{algorithm}[t]
\caption{EncryptedInference: Encrypted Linear Classifier Evaluation}
\label{alg:encrypted_inference}
\begin{algorithmic}[1]
\STATE \textbf{Input:} Encrypted replicated query $\overline{\mathbf{f}}_{\text{rep}}$, diagonal plaintexts $\{\text{diag}_d\}_{d=0}^{n_{\text{padded}}-1}$, bias plaintext $\text{bias}_{\text{pt}}$
\STATE \textbf{Output:} Encrypted results $\overline{\boldsymbol{\ell}}$
\STATE $\text{hoisted} \leftarrow \text{EvalFastRotationPrecompute}(\overline{\mathbf{f}}_{\text{rep}})$
\STATE $\overline{\text{acc}} \leftarrow \text{EvalMult}(\overline{\mathbf{f}}_{\text{rep}}, \text{diag}_0)$
\FOR{$d = 1$ \TO $n_{\text{padded}} - 1$}
	\STATE $\overline{\mathbf{f}}_{\text{rot}} \leftarrow \text{EvalFastRotation}(\overline{\mathbf{f}}_{\text{rep}}, d, \text{hoisted})$
	\STATE $\overline{\text{prod}} \leftarrow \text{EvalMult}(\overline{\mathbf{f}}_{\text{rot}}, \text{diag}_d)$
	\STATE $\overline{\text{acc}} \leftarrow \text{EvalAdd}(\overline{\text{acc}}, \overline{\text{prod}})$
\ENDFOR
\STATE \textcolor{blue}{$\text{elements\_per\_block} \leftarrow m_{\text{padded}}/n_{\text{padded}}$}
\STATE $\text{num\_shifts} \leftarrow \lceil\log_2(\text{elements\_per\_block})\rceil$
\STATE \textcolor{blue}{$\text{stride} \leftarrow n_{\text{padded}}$}
\FOR{$r = 0$ \TO $\text{num\_shifts}-1$}
	\STATE $\overline{\text{acc}} \leftarrow \text{EvalAdd}(\overline{\text{acc}}, \text{EvalRotate}(\overline{\text{acc}}, \text{stride}))$
	\STATE $\text{stride} \leftarrow 2\cdot\text{stride}$
\ENDFOR
\STATE $\overline{\boldsymbol{\ell}} \leftarrow \text{EvalAdd}(\overline{\text{acc}}, \text{bias}_{\text{pt}})$
\STATE \textbf{return} $\overline{\boldsymbol{\ell}}$
\end{algorithmic}
\end{algorithm}

To extract individual class scores from the decrypted packed vector, the edge applies the inverse of the RowHyDiag layout. Using the same block structure as bias encoding, for class index $i\in\{0,\ldots,n-1\}$ define
\begin{equation}
\text{blk}(i)=i\bmod n_{\text{pack}},\qquad
p(i)=\left\lfloor i/n_{\text{pack}}\right\rfloor,
\label{eq:decode_blk_p}
\end{equation}
and slot location
\begin{equation}
j(i)=\text{blk}(i)\,m_{\text{padded}}+p(i).
\label{eq:decode_slot_map}
\end{equation}
The recovered class score is then extracted from the corresponding slot:
\begin{equation}
\ell_i\approx\mathbf{a}_{\text{dec}}[j(i)],\quad i=0,\ldots,n-1.
\label{eq:result_index_map}
\end{equation}

Equivalently, if a slot is written as $j=\text{blk}\,m_{\text{padded}}+p$, then it stores class index $i=p\,n_{\text{pack}}+\text{blk}$ whenever $i<n$.

From~\eqref{eq:decode_blk_p}-\eqref{eq:result_index_map}, applying the slot mapping across valid classes yields
\begin{equation}
\boldsymbol{\ell}\approx\mathbf{W}\mathbf{f}+\mathbf{b},
\label{eq:equivalence_plaintext_encrypted}
\end{equation}
which shows that the three-stage trusted edge/untrusted cloud/trusted edge workflow preserves the plaintext linear-layer mapping, up to CKKS approximation error. Raw images, plaintext embeddings, and decryption keys remain inside the trusted edge boundary.

\section{Experiment Setup}
\label{sec:setup}
This section describes datasets, evaluation metrics, implementation settings, and baseline methods used to validate both prediction accuracy and deployment-oriented encrypted-inference performance.

\subsection{Datasets}
We evaluate the proposed framework on \textcolor{blue}{five} datasets spanning increasing levels of classification complexity and deployment realism: MNIST~\cite{lecun1998mnist} (10 classes), CIFAR-10~\cite{krizhevsky2009cifar} (10 classes), CIFAR-100~\cite{krizhevsky2009cifar} (100 classes), \textcolor{blue}{ImageNet-1K~\cite{deng2009imagenet} (1,000 classes),} and a custom Australian Landmarks dataset (9 classes).

MNIST is included as a standard low-complexity benchmark with grayscale handwritten digits and limited appearance variation, allowing us to verify numerical correctness and encrypted-plaintext consistency under controlled conditions. CIFAR-10 introduces natural-image diversity (background clutter, viewpoint changes, and intra-class diversity), which is useful for validating robustness beyond digit classification. CIFAR-100 further increases task difficulty through fine-grained 100-class recognition, providing a challenging evaluation of scalability for both model accuracy and encrypted inference overhead. \textcolor{blue}{ImageNet-1K extends the evaluation to 50,000 validation images and 1,000 output classes, allowing us to assess encrypted-plaintext consistency and encrypted-inference scalability at a substantially larger classification scale.}

In addition to standard benchmarks, we evaluate on a newly curated Australian landmarks dataset to reflect a realistic mobile visual-recognition scenario. This dataset contains 9 landmark classes and 1,981 test images used for final evaluation. We include this dataset because it captures deployment-oriented challenges (scene diversity, landmark-specific visual overlap, and edge-cloud deployment context) that are not fully represented by standard academic benchmarks. The dataset is constructed through web crawling followed by dual-stage filtering (semantic verification and visual consistency verification) to ensure data quality. For simplicity, only summary information is presented in the main text. The full data collection and verification pipeline is described in \appref{app:general_pipeline}, and Tables~\ref{tab:filtering_stats} and~\ref{tab:dataset_stats} in \appref{app:landmarks_stats} report the detailed per-class filtering statistics and final train/test split.

\subsection{Evaluation Metrics}

We evaluate three aspects of performance: classification quality, plaintext-encrypted numerical consistency, and system efficiency. For classification quality, top-1 accuracy is defined as
\begin{equation}
\text{Accuracy} = \frac{1}{N} \sum_{i=1}^{N} \mathbb{1}[\hat{y}_i = y_i],
\end{equation}
where $N$ is the number of test samples, $\hat{y}_i$ is the predicted class for sample $i$, $y_i$ is the true class label, and $\mathbb{1}[\cdot]$ is the indicator function. We also report confusion matrices for class-level analysis, where element $M_{ij}$ denotes the number of samples with true class $i$ predicted as class $j$. For system delay, we measure end-to-end latency as
\begin{equation}
T_{\text{total}} = T_{\text{edge}} + T_{\text{network}} + T_{\text{cloud}},
\end{equation}
where $T_{\text{edge}}$ includes both feature extraction time and encryption time, $T_{\text{network}}$ is network transmission time, and $T_{\text{cloud}}$ is homomorphic inference time. Throughput is reported in images per second and computed as $1/T_{\text{total}}$ when $T_{\text{total}}$ is measured in seconds (equivalently, $1000/T_{\text{total}}$ when latency is measured in milliseconds). In addition to runtime, we report ciphertext communication size and rotation-key storage to characterize deployment overhead. To quantify plaintext-encrypted numerical consistency, we compute the maximum relative error between plaintext logits $\mathbf{\ell} = [\ell_1, \ldots, \ell_C]$ and encrypted logits $\mathbf{\ell}' = [\ell'_1, \ldots, \ell'_C]$ across all test samples:
\begin{equation}
\text{Max-RelError} = \max_{i=1}^{N} \max_{c=1}^{C} \frac{|\ell_{i,c} - \ell'_{i,c}|}{|\ell_{i,c}| + \epsilon},
\end{equation}
where $\epsilon = 10^{-8}$ prevents division by zero, and $\ell_{i,c}$ denotes the logit for class $c$ of sample $i$.

\subsection{Experimental Configuration}

\subsubsection{Hardware and Network}
\begin{figure}[t]
\centering
\IfFileExists{images/Devices_Setup_v2.pdf}{
    \includegraphics[width=0.5\textwidth]{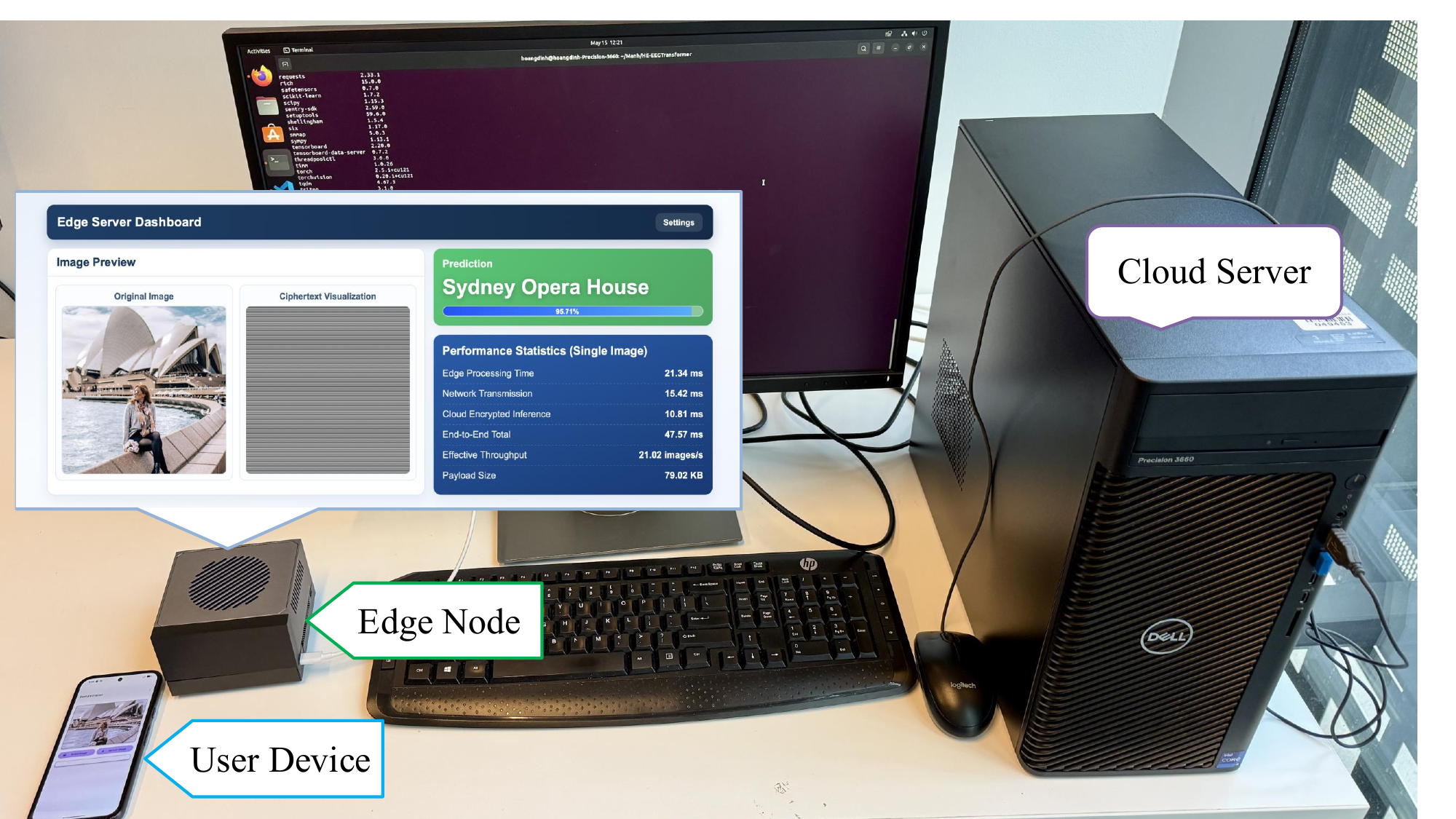}
}{
    \fbox{\parbox[c][3.2cm][c]{0.45\textwidth}{\centering Placeholder for real-device capture\\(Jetson Orin edge node, cloud server, and client device).}}
}
\caption{Real-device experimental setup used in the three-tier edge-cloud deployment.}
\label{fig:real_device_setup}
\end{figure}

Our implementation follows the proposed three-tier edge-cloud system shown in Fig.~\ref{fig:real_device_setup}, comprising an Android client, a trusted edge node, and an untrusted cloud server. The edge-to-cloud link uses 10 Gigabit Ethernet (10GbE). The trusted edge node is an NVIDIA Jetson Orin 64GB device, and the cloud server is equipped with an Intel Core i9-13900K processor (up to 5.8\,GHz, 24 cores / 32 threads) and 32\,GB of RAM running Linux. All latency components are measured empirically on this physical testbed using wall-clock timing. Specifically, $T_{\text{edge}}$ is measured end-to-end on the Jetson Orin, covering image preprocessing, feature extraction, and CKKS encryption; $T_{\text{network}}$ is obtained by timestamping ciphertext payloads at the sender and receiver endpoints, capturing the actual transfer delay over the 10GbE link; and $T_{\text{cloud}}$ is measured on the cloud server, covering the full homomorphic inference procedure (rotation, multiplication, accumulation, and bias addition). All measurements are taken under real operating conditions, and reported values are averaged over the full test set for each dataset.

\subsubsection{Model Backbones and Fine-Tuning Settings}
Model training and fine-tuning are conducted on the edge node to reflect deployment-oriented resource constraints, while encrypted classification is executed on the cloud. For MNIST, CIFAR-10, and Australian landmarks, we use a ResNet-34 backbone~\cite{he2016resnet} with 512-dimensional embeddings. For CIFAR-100, we use ViT-B/16~\cite{dosovitskiy2021vit} with 768-dimensional embeddings to support fine-grained 100-class recognition. \textcolor{blue}{For ImageNet-1K, we use a publicly available ViT-B/16 model pretrained on ImageNet-21K and fine-tuned on ImageNet-1K, also producing 768-dimensional embeddings for the $1{,}000$-class classifier.}

For ResNet-34 settings, we use Adam optimizer~\cite{kingma2015adam} with differential learning rates between the backbone and classifier, weight decay $5 \times 10^{-4}$, and validation-based checkpoint selection following standard transfer learning practices~\cite{kornblith2019better}. Specifically, MNIST uses 12 epochs with learning rates $1 \times 10^{-4}$ (backbone) and $1 \times 10^{-3}$ (classifier), CIFAR-10 uses 20 epochs with $5 \times 10^{-5}$ and $3 \times 10^{-4}$, and Australian landmarks uses 35 epochs with $5 \times 10^{-5}$ and $3 \times 10^{-4}$. The batch size is 64 in all three cases. For Australian landmarks, we additionally use a ReduceLROnPlateau scheduler~\cite{paszke2019pytorch} (factor 0.5, patience 3; minimum learning rate $10^{-7}$).

For CIFAR-100 with ViT-B/16, we fine-tune for 100 epochs using Adam (learning rate $7 \times 10^{-5}$, weight decay 0.05, batch size 128) following established ViT fine-tuning protocols~\cite{dosovitskiy2021vit,touvron2021deit}. We apply One-Cycle learning-rate scheduling~\cite{smith2019super}, Mixup ($\alpha=0.2$)~\cite{zhang2018mixup}, CutMix ($\alpha=1.0$)~\cite{yun2019cutmix}, RandomErasing~\cite{zhong2020random}, and label-smoothing cross-entropy (smoothing 0.1)~\cite{szegedy2016rethinking}, and select the best checkpoint by validation accuracy.

\subsubsection{HE Parameter Configuration}

{\color{blue}
\begin{table}[t]
\color{blue}
\centering
\caption{\textcolor{blue}{HE parameter sets used for encrypted inference.}}
\label{tab:ckks_params}
\renewcommand{\arraystretch}{1.1}
\setlength{\tabcolsep}{3.5pt}
\footnotesize
\begin{tabular}{lcccc}
\toprule
\textbf{Set} & $\mathbf{N}$ & \textbf{Modulus chain} & \textbf{Multiplicative Depth} & \textbf{Security Level} \\
\midrule
HE-2048  & 2,048  & $(60,50)$ & 1 & Prototype \\
HE-4096  & 4,096  & $(60,50)$ & 1 & Prototype \\
HE-8192  & 8,192  & $(60,50)$ & 1 & 128-bit \\
HE-16384 & 16,384 & $(60,50)$ & 1 & 128-bit \\
HE-32768 & 32,768 & $(60,50)$ & 1 & 128-bit \\
\bottomrule
\end{tabular}
\end{table}
}

{\color{blue}
The encrypted pipeline is implemented in OpenFHE~\cite{openfhe2022} using CKKS. Table~\ref{tab:ckks_params} defines five parameter sets, denoted by HE-2048 through HE-32768 according to their ring dimensions $N$; each set provides $S=N/2$ CKKS slots. All parameter sets use multiplicative depth one, a 60-bit first modulus, and a 50-bit scaling modulus. HE-8192, HE-16384, and HE-32768 are configured for 128-bit security, whereas HE-2048 and HE-4096 are included only as prototype parameter sets for evaluating computational scaling, without claiming a standard security level. The remaining implementation settings are \texttt{FIXEDAUTO} scaling, \texttt{HYBRID} key switching, and a uniform ternary secret distribution. These parameter choices follow the HomomorphicEncryption.org guidelines and OpenFHE's standard parameter interface~\cite{hestandard2018,hesecguidelines2024}.
}

\subsection{Baselines}

To comprehensively validate our proposed framework, we perform evaluations against baselines following standard practices in HE/FHE-based DL literature.

\textbf{Plaintext (Non-Encrypted) Baseline.} Similar to prior works~\cite{juvekar2018gazelle,cheon2025batch,manh2025iot}, we compare our approach against plaintext inference to quantify the impact of CKKS approximation noise. The plaintext baseline uses identical model weights and biases but evaluates the linear layer directly in floating-point arithmetic without encryption. This baseline is used to assess classification accuracy (top-1 accuracy, confusion matrices) and numerical consistency (maximum relative logit error between plaintext and encrypted outputs), verifying that encrypted inference preserves plaintext predictive behavior with negligible accuracy degradation.

\textbf{Halevi-Shoup Diagonal Encoding Baseline.} The Halevi-Shoup diagonal method~\cite{halevi2014algorithms} is a foundational technique for encrypted matrix-vector multiplication that decomposes a square matrix into diagonals, encodes them as plaintexts, and multiplies them with rotated encrypted vectors. Prior HE-based neural network inference approaches have widely adopted this method for encrypted linear operation~\cite{jiang2018secure,miran2022secure}, making it a natural baseline for evaluating the algorithmic contribution of RowHyDiag. The method embeds an $n \times m$ weight matrix into an $L \times L$ square matrix (where $L = \max(n, m)$ rounded to the next power of two), then extracts $L$ diagonal plaintexts for encrypted matrix-vector computation through rotation-and-multiply operations. The rotation count for Halevi-Shoup encoding is given in~\eqref{eq:rot_halevi_shoup}. This baseline enables direct comparison of cloud inference efficiency metrics (latency and memory) to demonstrate that RowHyDiag significantly reduces computational overhead, memory footprint, and inference latency compared to conventional diagonal encoding.

{\color{blue}
\textbf{PEGASUS Linear-Transformation Baseline.} PEGASUS~\cite{lu2021pegasus} uses a tiled baby-step/giant-step (BSGS) method for encrypted linear transformations. We adapt this method to evaluate the encrypted classifier $\boldsymbol{\ell}=W\mathbf{f}+\mathbf{b}$. The method pads the matrix dimensions to powers of two and constructs $D=\min(n_{\mathrm{padded}},m_{\mathrm{padded}})$ tiled diagonals. It then applies a balanced BSGS schedule with baby-step size $\lceil\sqrt{D}\rceil$. Because our framework performs only encrypted linear-layer evaluation at the cloud, the comparison does not include PEGASUS's conversion, lookup-table evaluation, or scheme-switching operations. Therefore, this baseline represents only the linear-transformation method in PEGASUS, rather than the complete PEGASUS system.

\textbf{RowHyDiag Evaluation Variants.} RowHyDiag is designed primarily for short-wide classifier matrices, where the number of output classes is much smaller than the feature dimension. However, as the number of output classes increases, the number of diagonal products also grows. To reduce the resulting rotation overhead, we apply BSGS scheduling, which changes only the order in which the diagonal products are accumulated. Accordingly, we use RowHyDiag for the three small-output datasets and RowHyDiag with BSGS scheduling for CIFAR-100 and ImageNet-1K.
}

\section{Simulation and Experimental Results}
\label{sec:results}

Before evaluating encrypted inference, we verify that all fine-tuned models converge to stable operating points with small train-validation gaps. Convergence curves and detailed analysis for \textcolor{blue}{MNIST, CIFAR-10, CIFAR-100, and Australian landmarks} are provided in Fig.~\ref{fig:training_convergence_all} in \appref{app:convergence}. The selected checkpoints achieve 99.47\% (MNIST), 97.05\% (CIFAR-10), 92.69\% (CIFAR-100), and 93.92\% (Australian landmarks) validation accuracy, confirming reliable feature representations for downstream encrypted classification. \textcolor{blue}{For ImageNet-1K, we instead use the public ViT-B/16 AugReg checkpoint distributed through the \texttt{timm} model collection on Hugging Face.\footnote{\url{https://huggingface.co/timm/vit_base_patch16_384.augreg_in21k_ft_in1k}}}

\subsection{Encrypted Classification Accuracy}

\begin{figure*}[t]
\centering
\begin{subfigure}[b]{0.32\textwidth}
\centering
\includegraphics[width=\linewidth]{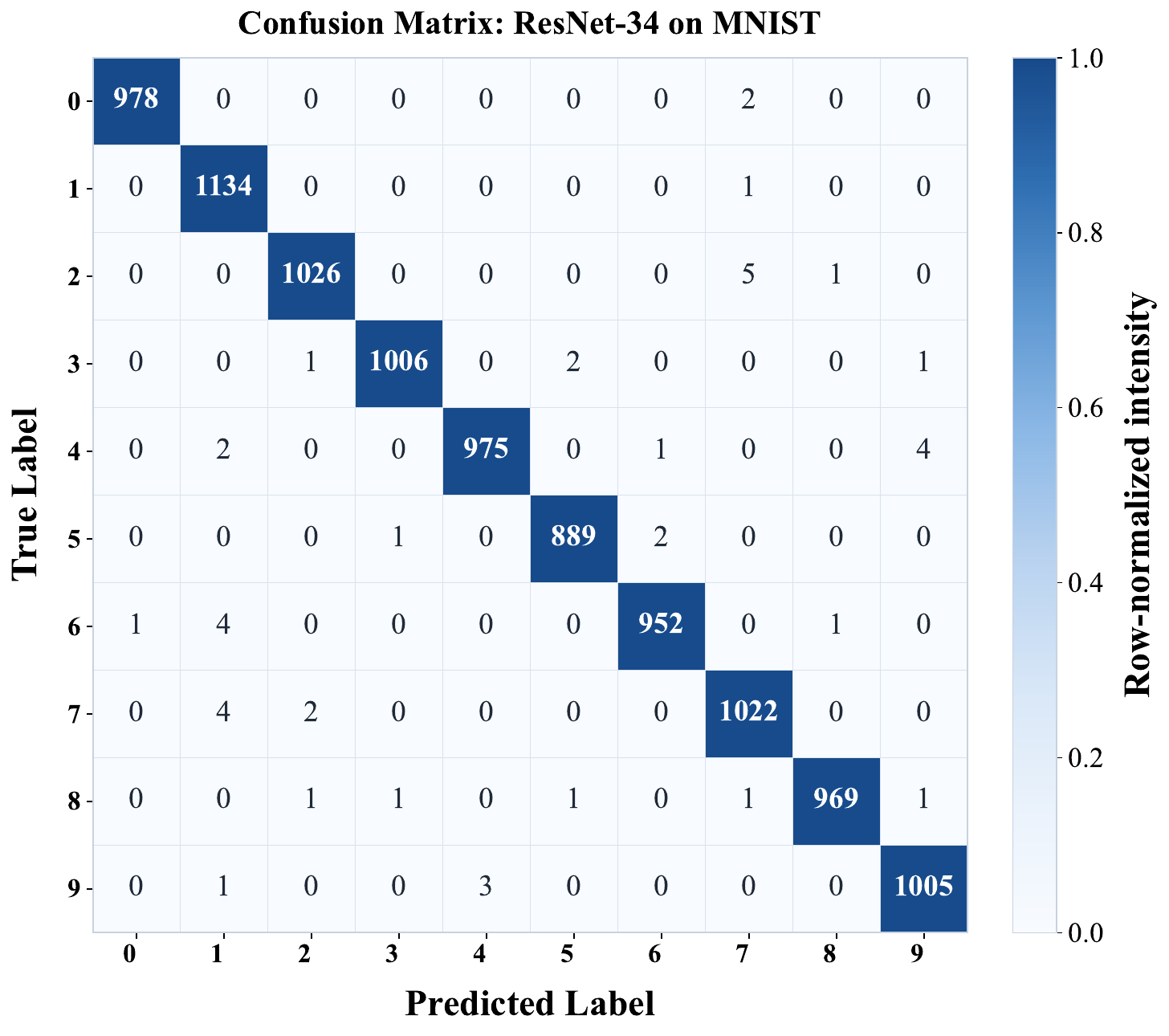}
\caption{MNIST (10 classes).}
\label{fig:enc_conf_mnist}
\end{subfigure}
\hfill
\begin{subfigure}[b]{0.32\textwidth}
\centering
\includegraphics[width=\linewidth]{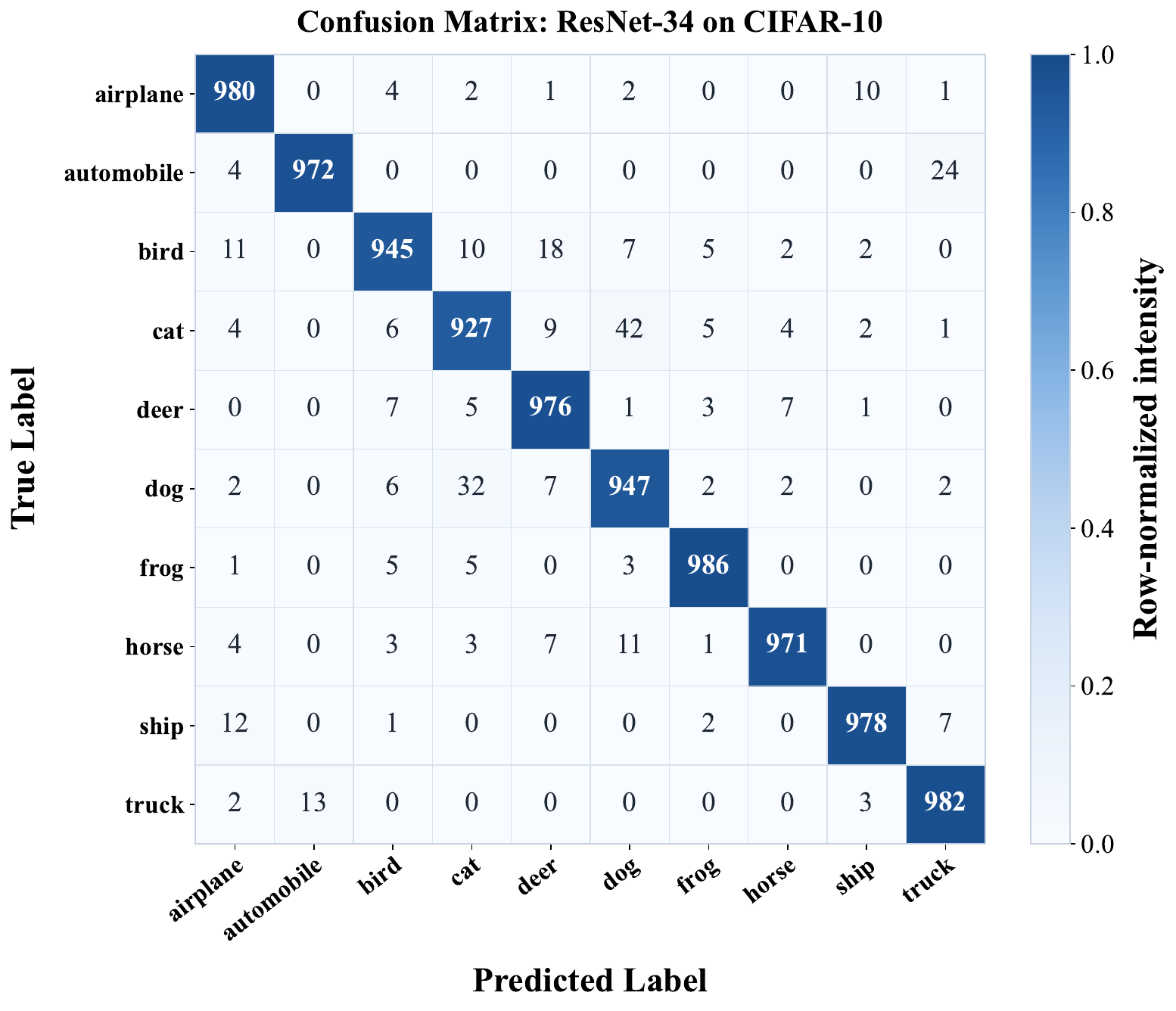}
\caption{CIFAR-10 (10 classes).}
\label{fig:enc_conf_cifar10}
\end{subfigure}
\hfill
\begin{subfigure}[b]{0.32\textwidth}
\centering
\includegraphics[width=\linewidth]{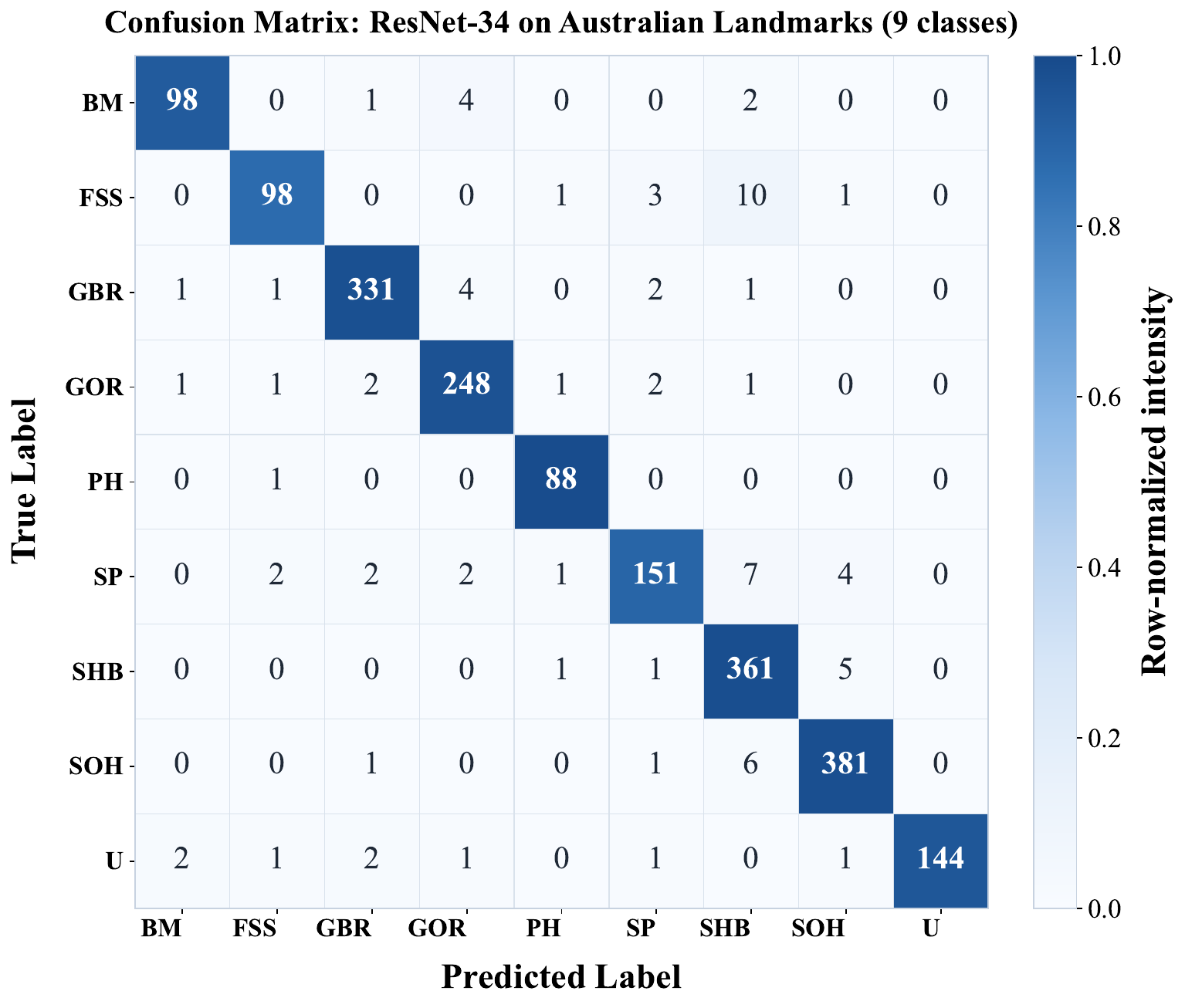}
\caption{Australian landmarks (9 classes).}
\label{fig:enc_conf_landmarks}
\end{subfigure}
\caption{\textcolor{blue}{Encrypted confusion matrices for the three 9/10-class datasets. The CIFAR-100 and ImageNet-1K matrices are not shown because their 100 and 1,000 classes make them too dense for clear visualization in the main text.}}
\label{fig:enc_confusion_all}
\end{figure*}

{\color{blue}
Table~\ref{tab:enc_acc_summary} summarizes plaintext-versus-encrypted accuracy and the maximum relative logit error for every evaluated HE parameter set. Encrypted and plaintext top-1 accuracies are identical for all five datasets and all tested configurations. The largest observed relative error is $3.216\times10^{-4}$ on ImageNet-1K under HE-32768, which remains too small to alter any top-1 decision. The errors across the remaining parameter sets are of the same negligible order and likewise preserve all top-1 decisions.

\begin{table}[t]
\color{blue}
\centering
\caption{\textcolor{blue}{Encrypted classification accuracy and maximum relative logit error across HE parameter sets.}}
\label{tab:enc_acc_summary}
\renewcommand{\arraystretch}{1.08}
\setlength{\tabcolsep}{2.5pt}
\footnotesize
\begin{tabular}{@{}llccc@{}}
\toprule
\textbf{Dataset} & \textbf{Set} & \textbf{Plain (\%)} & \textbf{Encrypted (\%)} & \textbf{Max-Rel. Error} \\
\midrule
\multirow{3}{*}{MNIST} & HE-2048 & \multirow{3}{*}{99.56} & \multirow{3}{*}{99.56} & $2.659\times10^{-8}$ \\
 & HE-4096 & & & $3.510\times10^{-7}$ \\
 & HE-8192 & & & $2.208\times10^{-7}$ \\
\midrule
\multirow{3}{*}{CIFAR-10} & HE-2048 & \multirow{3}{*}{96.64} & \multirow{3}{*}{96.64} & $9.671\times10^{-9}$ \\
 & HE-4096 & & & $2.400\times10^{-8}$ \\
 & HE-8192 & & & $2.956\times10^{-8}$ \\
\midrule
\multirow{3}{*}{Landmarks} & HE-2048 & \multirow{3}{*}{95.91} & \multirow{3}{*}{95.91} & $1.472\times10^{-8}$ \\
 & HE-4096 & & & $1.567\times10^{-8}$ \\
 & HE-8192 & & & $2.196\times10^{-8}$ \\
\midrule
\multirow{4}{*}{CIFAR-100} & HE-4096 & \multirow{4}{*}{92.69} & \multirow{4}{*}{92.69} & $1.053\times10^{-5}$ \\
 & HE-8192 & & & $3.666\times10^{-5}$ \\
 & HE-16384 & & & $1.153\times10^{-4}$ \\
 & HE-32768 & & & $1.844\times10^{-4}$ \\
\midrule
\multirow{4}{*}{ImageNet-1K} & HE-4096 & \multirow{4}{*}{85.99} & \multirow{4}{*}{85.99} & $7.096\times10^{-5}$ \\
 & HE-8192 & & & $3.080\times10^{-4}$ \\
 & HE-16384 & & & $9.484\times10^{-5}$ \\
 & HE-32768 & & & $3.216\times10^{-4}$ \\
\bottomrule
\end{tabular}
\end{table}
}

\textcolor{blue}{To provide class-level transparency for the three 9/10-class datasets, Fig.~\ref{fig:enc_confusion_all} combines the encrypted confusion matrices for MNIST, CIFAR-10, and Australian landmarks. The matrices show that prediction errors are concentrated in visually similar classes rather than randomly distributed. Together with the identical plaintext and encrypted accuracies for CIFAR-100 and ImageNet-1K in Table~\ref{tab:enc_acc_summary}, these results show that the packing and reduction procedure preserves model utility from 9 to 1,000 output classes.}

\subsection{Low-Latency Encrypted Inference Analysis}

\subsubsection{Latency and Throughput Analysis}

\begin{figure}[t]
\centering
\includegraphics[width=\columnwidth]{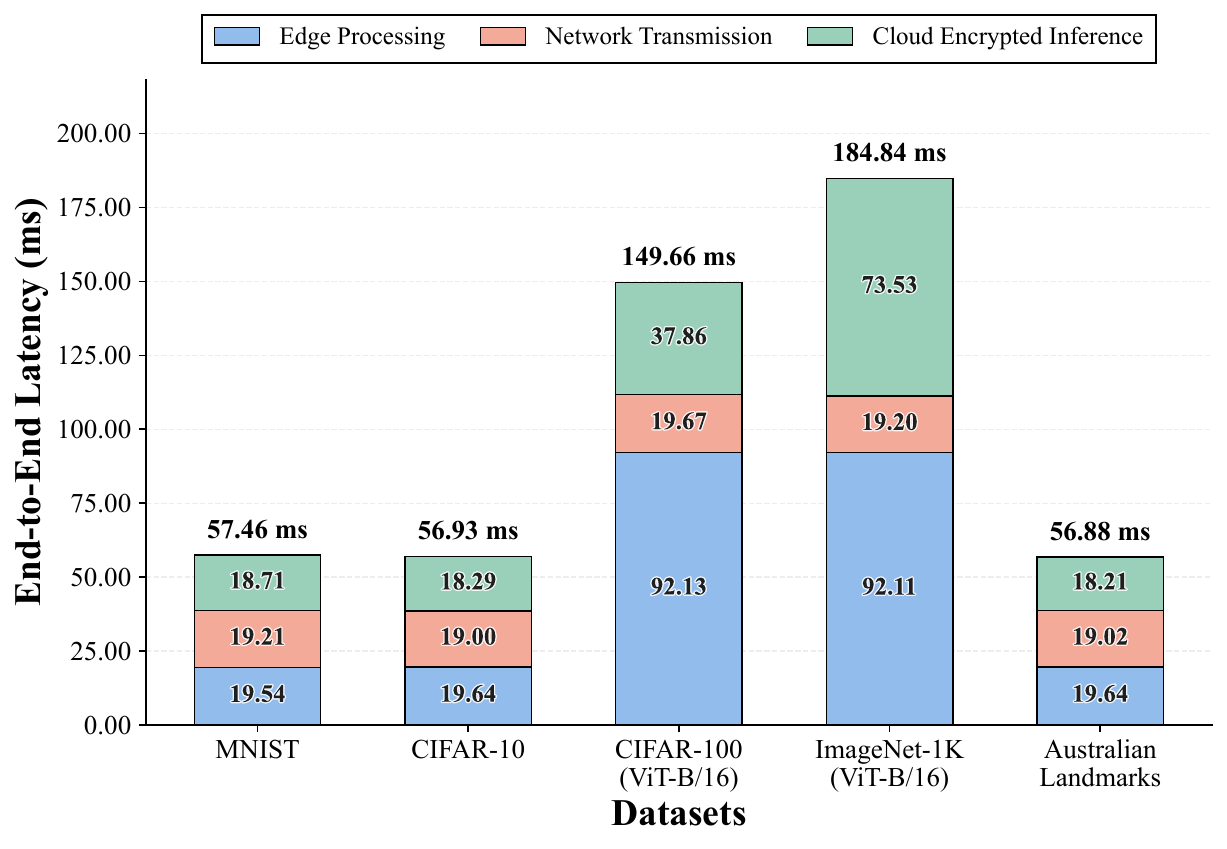}
\caption{Per-stage latency breakdown across datasets \textcolor{blue}{under the HE-8192 parameter set}. Stacked bars show edge processing (blue), network transmission (red), and cloud encrypted inference (green). Total latency is annotated above each bar.}
\label{fig:latency_breakdown}
\end{figure}

\textcolor{blue}{Fig.~\ref{fig:latency_breakdown} presents the end-to-end latency breakdown across all five datasets under the common HE-8192 parameter set.}

{\color{blue}
For MNIST, CIFAR-10, and Australian Landmarks, the end-to-end latencies are 57.46, 56.93, and 56.88\,ms, corresponding to throughputs of 17.40, 17.57, and 17.58 images/s, respectively. The three latency components are nearly balanced in these settings: edge processing requires approximately 19.54-19.64\,ms, network transmission requires 19.00-19.21\,ms, and cloud encrypted inference requires 18.21-18.71\,ms. These results show that the proposed pipeline supports interactive encrypted classification for the 9/10-class workloads, with no individual stage becoming a dominant bottleneck.

The two larger ViT-B/16 workloads exhibit the expected increase in latency. CIFAR-100 requires 92.13\,ms for edge processing, 19.67\,ms for network transmission, and 37.86\,ms for cloud encrypted inference, resulting in 149.66\,ms end-to-end latency and 6.68 images/s throughput. ImageNet-1K requires 92.11, 19.20, and 73.53\,ms for the corresponding stages, yielding 184.84\,ms total latency and 5.41 images/s. The similar edge-processing times follow from the common ViT-B/16 backbone and 768-dimensional feature representation. Their network times are also similar because both datasets use the HE-8192 parameter set. In contrast, the cloud component increases from 37.86\,ms for CIFAR-100 to 73.53\,ms for ImageNet-1K because the encrypted classifier scales from 100 to 1,000 output classes. Overall, cloud inference latency increases with the number of output classes, while end-to-end latency remains below 200\,ms for all evaluated workloads.
}

\subsubsection{Complexity Analysis of Rotation Count}

To understand the algorithmic efficiency of the proposed encoding, we analyze the theoretical rotation complexity of the conventional Halevi-Shoup diagonal method~\cite{halevi2014algorithms}, the PEGASUS linear-transformation baseline, and RowHyDiag. Rotation operations (via key-switching) are a major computational cost in encrypted linear-layer evaluation, making rotation count an important complexity metric for encrypted inference systems~\cite{jiang2018secure}.

\textbf{Halevi-Shoup Diagonal Encoding.} Consider an encrypted matrix-vector computation $\mathbf{y} = \mathbf{A} \mathbf{x}$ where $\mathbf{A} \in \mathbb{R}^{n \times m}$ is a plaintext matrix ($n$: number of output rows, $m$: number of input columns) and $\mathbf{x} \in \mathbb{R}^{m}$ is encrypted. The Halevi-Shoup diagonal method~\cite{halevi2014algorithms} embeds $\mathbf{A}$ into a square matrix of size $L \times L$ where $L = 2^{\lceil \log_2 \max(n,m) \rceil}$ is the smallest power of two greater than or equal to $\max(n,m)$. The method then extracts $L$ diagonal plaintexts and performs encrypted matrix-vector multiplication through rotation-and-multiply operations. For each diagonal $d \in \{0, 1, \ldots, L-1\}$, the encrypted vector is rotated by $d$ positions, multiplied with the corresponding diagonal plaintext, and the resulting $L$ ciphertexts are summed via homomorphic addition. Since no further reduction is needed (each output element is already in its final position after summation), the total rotation count is~\cite{halevi2014algorithms}:
\begin{equation}
R_{\text{Halevi-Shoup}}(n, m) = L - 1 = 2^{\lceil \log_2 \max(n,m) \rceil} - 1
\label{eq:rot_halevi_shoup}
\end{equation}
For typical classification tasks where $m \gg n$, the rotation count is dominated by the input dimension and scales as $O(m)$.

\textcolor{blue}{\textbf{PEGASUS Linear-Transformation.} For output dimension $n$ and input dimension $m$, the baseline pads the two dimensions to $\widehat n=2^{\lceil\log_2n\rceil}$ and $\widehat m=2^{\lceil\log_2m\rceil}$, respectively. The resulting tiled representation contains}
\begin{equation}
\textcolor{blue}{D_P=\min(\widehat n,\widehat m)}.
\label{eq:pegasus_diagonals}
\end{equation}
\textcolor{blue}{The baseline evaluates these diagonals using a balanced BSGS schedule with baby-step size $B_P=\lceil\sqrt{D_P}\rceil$ and $G_P=\lceil D_P/B_P\rceil$ giant groups. Preparing the reusable baby steps requires $B_P-1$ rotations, while accumulating the giant groups requires $G_P-1$ rotations. When $\widehat n<\widehat m$, each output is distributed across $\widehat m/\widehat n$ column groups. Combining these partial values requires an additional $\log_2(\widehat m/\widehat n)$ rotations. Substituting the padded dimensions and BSGS parameters gives the following rotation count in terms of $n$ and $m$:}
{\color{blue}
\begin{equation}
\resizebox{0.96\columnwidth}{!}{$
\begin{aligned}
R_{\text{PEGASUS LT}}(n,m)
={}&\left\lceil\sqrt{\min\!\left(2^{\lceil\log_2n\rceil},2^{\lceil\log_2m\rceil}\right)}\right\rceil-1\\
&+\left\lceil\frac{\min\!\left(2^{\lceil\log_2n\rceil},2^{\lceil\log_2m\rceil}\right)}
{\left\lceil\sqrt{\min\!\left(2^{\lceil\log_2n\rceil},2^{\lceil\log_2m\rceil}\right)}\right\rceil}\right\rceil-1\\
&+\max\!\left(0,\log_2\frac{2^{\lceil\log_2m\rceil}}{2^{\lceil\log_2n\rceil}}\right).
\end{aligned}
$}
\label{eq:rot_pegasus_lt}
\end{equation}
}

\textbf{RowHyDiag Encoding.} In contrast, RowHyDiag uses the available CKKS slots to pack multiple classifier rows before extracting the diagonals. For slot count $S=N/2$ and input dimension $m$, the feature length is first padded to
\begin{equation}
m_{\mathrm{padded}}=2^{\lceil\log_2m\rceil}.
\label{eq:m_padded_complexity}
\end{equation}
The number of classifier rows combined in each packed row is
\begin{equation}
n_{\mathrm{pack}}=\left\lfloor\frac{S}{2^{\lceil\log_2m\rceil}}\right\rfloor.
\label{eq:n_pack}
\end{equation}
The resulting number of packed rows is then padded to a power of two:
\begin{equation}
n_{\mathrm{padded}}
=2^{\left\lceil\log_2\left\lceil\frac{n}{n_{\mathrm{pack}}}\right\rceil\right\rceil}.
\label{eq:n_padded_complexity}
\end{equation}
The direct schedule evaluates the $n_{\mathrm{padded}}$ diagonal products using $n_{\mathrm{padded}}-1$ hoisted rotations. \textcolor{blue}{It then combines the partial feature contributions through an in-block binary-tree reduction, which requires $\lceil\log_2(m_{\mathrm{padded}}/n_{\mathrm{padded}})\rceil$ additional rotations.} Substituting Eqs.~\eqref{eq:m_padded_complexity}-\eqref{eq:n_padded_complexity} gives the following rotation count in terms of $n$, $m$, and $S$:
\begin{equation}
\resizebox{0.96\columnwidth}{!}{$
\begin{aligned}
R_{\text{RowHyDiag}}(n,m,S)
={}&2^{\left\lceil\log_2\left\lceil
\frac{n}{\left\lfloor S/2^{\lceil\log_2m\rceil}\right\rfloor}
\right\rceil\right\rceil}-1\\
&+\textcolor{blue}{\left\lceil\log_2\!\left(
\frac{2^{\lceil\log_2m\rceil}}
{2^{\left\lceil\log_2\left\lceil n/\left\lfloor S/2^{\lceil\log_2m\rceil}\right\rfloor\right\rceil\right\rceil}}
\right)\right\rceil}.
\end{aligned}
$}
\label{eq:rot_rowpacked}
\end{equation}

\textcolor{blue}{For MNIST, CIFAR-10, and Australian Landmarks, row packing produces a small number of diagonals, so we use the direct schedule. For the larger-output CIFAR-100 and ImageNet-1K classifiers, we instead apply BSGS to the same $n_{\mathrm{padded}}$ diagonal products. We use baby-step size $B_R=\lceil\sqrt{n_{\mathrm{padded}}}\rceil$ and $G_R=\lceil n_{\mathrm{padded}}/B_R\rceil$ giant groups. This reduces the rotations in the diagonal stage from $n_{\mathrm{padded}}-1$ to $(B_R-1)+(G_R-1)$, while the row-packed encoding and in-block reduction remain unchanged. Substituting the padded dimensions gives the following rotation count in terms of $n$, $m$, and $S$:}
{\color{blue}
\begin{equation}
\resizebox{0.96\columnwidth}{!}{$
\begin{aligned}
R_{\text{RowHyDiag}}^{\mathrm{BSGS}}(n,m,S)
={}&\left\lceil\sqrt{2^{\left\lceil\log_2\left\lceil n/\left\lfloor S/2^{\lceil\log_2m\rceil}\right\rfloor\right\rceil\right\rceil}}\right\rceil-1\\
&+\left\lceil\frac{2^{\left\lceil\log_2\left\lceil n/\left\lfloor S/2^{\lceil\log_2m\rceil}\right\rfloor\right\rceil\right\rceil}}
{\left\lceil\sqrt{2^{\left\lceil\log_2\left\lceil n/\left\lfloor S/2^{\lceil\log_2m\rceil}\right\rfloor\right\rceil\right\rceil}}\right\rceil}\right\rceil-1\\
&+\left\lceil\log_2\!\left(
\frac{2^{\lceil\log_2m\rceil}}
{2^{\left\lceil\log_2\left\lceil n/\left\lfloor S/2^{\lceil\log_2m\rceil}\right\rfloor\right\rceil\right\rceil}}
\right)\right\rceil.
\end{aligned}
$}
\label{eq:rot_rowpacked_bsgs}
\end{equation}
}

\textcolor{blue}{Table~\ref{tab:rotation_complexity} reports the rotation counts obtained from Eqs.~\eqref{eq:rot_halevi_shoup}, \eqref{eq:rot_pegasus_lt}, \eqref{eq:rot_rowpacked}, and~\eqref{eq:rot_rowpacked_bsgs}.}

\begin{table*}[t]
\color{blue}
\centering
\caption{\textcolor{blue}{Rotation counts across the evaluated slot counts. The smallest count in each setting is shown in bold.}}
\label{tab:rotation_complexity}
\renewcommand{\arraystretch}{1.08}
\setlength{\tabcolsep}{9pt}
\footnotesize
\begin{tabular}{@{}llrrrr@{}}
\toprule
\textbf{Dataset} & $\boldsymbol{n\times m}$ & $\boldsymbol{S}$ & \textbf{Halevi-Shoup} & \textbf{PEGASUS LT} & \textbf{RowHyDiag} \\
\midrule
\multirow{3}{*}{\shortstack[l]{MNIST, CIFAR-10,\\Australian Landmarks}}
 & \multirow{3}{*}{$(9\text{-}10)\times512$} & 1,024 & 511 & \textbf{11} & 13 \\
 & & 2,048 & 511 & 11 & \textbf{10} \\
 & & 4,096 & 511 & 11 & \textbf{9} \\
\midrule
\multirow{4}{*}{CIFAR-100}
 & \multirow{4}{*}{$100\times768$} & 2,048  & 1,023 & 24 & \textbf{18} \\
 & & 4,096  & 1,023 & 24 & \textbf{15} \\
 & & 8,192  & 1,023 & 24 & \textbf{12} \\
 & & 16,384 & 1,023 & 24 & \textbf{11} \\
\midrule
\multirow{4}{*}{ImageNet-1K}
 & \multirow{4}{*}{$1{,}000\times768$} & 2,048  & 1,023 & 62 & \textbf{45} \\
 & & 4,096  & 1,023 & 62 & \textbf{32} \\
 & & 8,192  & 1,023 & 62 & \textbf{24} \\
 & & 16,384 & 1,023 & 62 & \textbf{18} \\
\bottomrule
\end{tabular}
\end{table*}

\textcolor{blue}{\textbf{Comparison.} For the 9/10-class tasks, RowHyDiag decreases from 13 to 9 rotations as $S$ increases from 1,024 to 4,096, while Halevi-Shoup and the PEGASUS linear-transformation baseline remain at 511 and 11 rotations, respectively. For CIFAR-100, RowHyDiag with BSGS decreases from 18 to 11 rotations over $S=2{,}048$-$16{,}384$, compared with 24 for the PEGASUS baseline and 1,023 for Halevi-Shoup. For ImageNet-1K, the RowHyDiag count decreases from 45 to 18, compared with 62 and 1,023 rotations for the two baselines. These reductions result from slot-aware row packing, which packs more output rows together as $S$ increases and therefore reduces the number of RowHyDiag diagonals.}

\textcolor{blue}{These results are consistent with the asymptotic rotation complexity. Power-of-two padding changes each dimension by at most a constant factor. Therefore, Halevi-Shoup uses $\Theta(\max(n,m))$ diagonals and has rotation complexity $O(\max(n,m))$. The PEGASUS linear-transformation baseline uses $D_P=\Theta(\min(n,m))$ tiled diagonals. Its balanced BSGS schedule therefore has rotation complexity}
{\color{blue}
\begin{equation}
O\!\left(\sqrt{\min(n,m)}+\max\!\left(0,\log\frac{m}{n}\right)\right).
\label{eq:complexity_pegasus}
\end{equation}
}
\textcolor{blue}{For RowHyDiag, $n_{\mathrm{pack}}=\Theta(S/m)$ and hence the number of packed diagonals is}
{\color{blue}
\begin{equation}
D_R=n_{\mathrm{padded}}=\Theta\!\left(\left\lceil\frac{nm}{S}\right\rceil\right),
\label{eq:rowhydiag_diagonal_complexity}
\end{equation}
}
\textcolor{blue}{assuming a valid packing configuration with $S\ge 2^{\lceil\log_2m\rceil}$. The direct RowHyDiag schedule thus has rotation complexity}
{\color{blue}
\begin{equation}
O\!\left(D_R+\log\frac{m}{D_R}\right),
\label{eq:complexity_rowhydiag_direct}
\end{equation}
}
\textcolor{blue}{whereas applying BSGS to its diagonal stage reduces the rotation component to}
{\color{blue}
\begin{equation}
O\!\left(\sqrt{D_R}+\log\frac{m}{D_R}\right)
=O\!\left(\sqrt{\left\lceil\frac{nm}{S}\right\rceil}
+\log\frac{m}{\left\lceil nm/S\right\rceil}\right).
\label{eq:complexity_rowhydiag_bsgs}
\end{equation}
}
\textcolor{blue}{In the intended short-wide regime, where $n$ is small and $S=\Theta(m)$, $D_R$ remains small and the RowHyDiag rotation complexity approaches $O(\log m)$. BSGS changes the rotation schedule but does not reduce the number of ciphertext-plaintext products. RowHyDiag evaluates $D_R$ diagonal products, whereas the PEGASUS linear-transformation baseline evaluates $D_P$ products. Their diagonal-product counts are therefore $O(D_R)$ and $O(D_P)$, respectively. Thus, rotation count captures the main key-switching cost but does not alone determine the complete homomorphic runtime.}

\subsubsection{Memory and Resource Analysis}

Beyond computational complexity, deployable systems require careful consideration of \textcolor{blue}{storage} and communication overhead. \textcolor{blue}{We therefore report two deployment-oriented metrics: the one-time setup payload and per-query communication volume. The setup payload includes the cryptographic context, public key, and rotation keys provided to and stored by the cloud. Per-query communication includes the uploaded feature ciphertext and returned result ciphertext. Table~\ref{tab:memory_resource_summary} summarizes these costs under the common HE-8192 parameter set.}

\begin{table}[t]
\color{blue}
\centering
\caption{\textcolor{blue}{Memory and resource summary for RowHyDiag using the HE-8192 parameter set (decimal MB/KB).}}
\label{tab:memory_resource_summary}
\resizebox{\columnwidth}{!}{%
\begin{tabular}{lccc}
\toprule
\textbf{Metric} & \textcolor{blue}{\textbf{9/10 classes}} & \textbf{CIFAR-100} & \textcolor{blue}{\textbf{ImageNet-1K}} \\
\midrule
Setup (MB) & \textcolor{blue}{7.35} & \textcolor{blue}{12.07} & \textcolor{blue}{25.45} \\
Context+PK (KB) & \textcolor{blue}{263.54} & \textcolor{blue}{263.54} & \textcolor{blue}{263.54} \\
Rotation keys (MB) & \textcolor{blue}{7.08} & \textcolor{blue}{11.81} & \textcolor{blue}{25.18} \\
\textcolor{blue}{Rotation indices (\#)} & \textcolor{blue}{9} & \textcolor{blue}{15} & \textcolor{blue}{32} \\
\textcolor{blue}{Upload/query (KB)} & \textcolor{blue}{263.01} & \textcolor{blue}{263.01} & \textcolor{blue}{263.01} \\
\textcolor{blue}{Return/query (KB)} & \textcolor{blue}{263.21} & \textcolor{blue}{263.21} & \textcolor{blue}{263.21} \\
\bottomrule
\end{tabular}
}
\end{table}

\textcolor{blue}{For the 9/10-class settings (MNIST, CIFAR-10, and Australian Landmarks), the setup payload is 7.35\,MB, including 263.54\,KB for the context and public key and 7.08\,MB for nine rotation keys. Rotation keys account for approximately 96.3\% of the setup payload, indicating that evaluation-key material, rather than the query ciphertext, is the dominant static overhead. At inference time, the encrypted upload and returned result are 263.01\,KB and 263.21\,KB per image, respectively. The near-equal upload and return sizes follow from the slot-position-matching decode strategy: the cloud returns one logits ciphertext in the same packed layout without ciphertext expansion. These values are shared by the three datasets because they use the same HE-8192 configuration and comparable packing layouts.}

\textcolor{blue}{For CIFAR-100, the larger output dimension increases the number of required rotation indices from 9 to 15. Consequently, rotation-key storage grows to 11.81\,MB and the total setup payload reaches 12.07\,MB. ImageNet-1K further increases the output dimension to 1,000 classes, requiring 32 rotation indices, 25.18\,MB of rotation keys, and a 25.45\,MB total setup payload. Rotation keys therefore remain the dominant setup component, accounting for approximately 97.8\% and 98.9\% of the CIFAR-100 and ImageNet-1K setup payloads, respectively.}

\textcolor{blue}{Because all three settings use the same ring dimension, their per-query upload and return sizes remain unchanged; the larger class counts primarily affect the number of required rotation keys. Relative to the 9/10-class settings, setup payload increases by approximately 1.6$\times$ for CIFAR-100 and 3.5$\times$ for ImageNet-1K. At the 1,000-class scale, the RowHyDiag setup payload is 25.45\,MB.}

\subsubsection{\textcolor{blue}{Empirical Comparison with Baselines}}

{\color{blue}
Having established the theoretical rotation advantage, we compare our proposed approach RowHyDiag with the conventional Halevi-Shoup diagonal method and the PEGASUS linear-transformation baseline. All methods use the same model, cryptographic parameters, hardware, and evaluation settings. Table~\ref{tab:empirical_comparison} reports cloud inference latency and one-time setup payload. The setup payload includes the cryptographic context, public key, and rotation keys.

\begin{table}[t]
\color{blue}
\centering
\caption{Empirical performance comparison of RowHyDiag and baselines. Entries report cloud latency (ms/image) / setup payload (MB).}
\label{tab:empirical_comparison}
\scriptsize
\setlength{\tabcolsep}{2.5pt}
\renewcommand{\arraystretch}{1.03}
\resizebox{\columnwidth}{!}{%
\begin{tabular}{llrrr}
\toprule
\textbf{Dataset} & \textbf{Set} & \textbf{Halevi-Shoup} & \textbf{PEGASUS LT} & \textbf{RowHyDiag} \\
\midrule
\multirow{3}{*}{MNIST}
 & HE-2048 & 167.79 / 100.82 & 6.79 / \textbf{2.24} & \textbf{6.29} / 2.63 \\
 & HE-4096 & 253.50 / 201.35 & 12.07 / 4.46 & \textbf{10.71 / 4.07} \\
 & HE-8192 & 418.57 / 402.41 & 19.79 / 8.92 & \textbf{18.64 / 7.35} \\
\midrule
\multirow{3}{*}{CIFAR-10}
 & HE-2048 & 168.00 / 100.82 & 7.14 / \textbf{2.24} & \textbf{6.86} / 2.63 \\
 & HE-4096 & 253.07 / 201.35 & 11.07 / 4.46 & \textbf{8.21 / 4.07} \\
 & HE-8192 & 418.57 / 402.41 & 19.71 / 8.92 & \textbf{18.07 / 7.35} \\
\midrule
\multirow{4}{*}{CIFAR-100}
 & HE-4096 & 554.57 / 402.96 & 36.64 / 9.58 & \textbf{31.29 / 7.22} \\
 & HE-8192 & 899.93 / 805.35 & 55.00 / 19.15 & \textbf{37.86 / 12.07} \\
 & HE-16384 & 1629.00 / 1610.13 & 85.86 / 38.29 & \textbf{51.21 / 19.41} \\
 & HE-32768 & - & 153.21 / 76.56 & \textbf{83.07 / 35.66} \\
\midrule
\multirow{4}{*}{ImageNet-1K}
 & HE-4096 & 557.21 / 402.96 & 117.74 / 24.55 & \textbf{68.11 / 17.85} \\
 & HE-8192 & 895.53 / 805.35 & 180.21 / 49.06 & \textbf{73.53 / 25.45} \\
 & HE-16384 & 1610.89 / 1610.13 & 306.11 / 98.08 & \textbf{89.89 / 38.29} \\
 & HE-32768 & - & 536.79 / 196.12 & \textbf{119.42 / 57.68} \\
\midrule
\multirow{3}{*}{Australian Landmarks}
 & HE-2048 & 166.93 / 100.82 & 7.36 / \textbf{2.24} & \textbf{7.21} / 2.63 \\
 & HE-4096 & 251.50 / 201.35 & 11.64 / 4.46 & \textbf{10.50 / 4.07} \\
 & HE-8192 & 418.00 / 402.41 & 19.00 / 8.92 & \textbf{18.07 / 7.35} \\
\bottomrule
\end{tabular}
}
\end{table}

\textbf{Comparison with Halevi-Shoup.} Under the prototype HE-2048 and HE-4096 settings, our proposed approach RowHyDiag is 8.2-30.8$\times$ faster and reduces the setup payload by 22.6-55.8$\times$. Under the configurations providing 128-bit security, RowHyDiag is 12.2-31.8$\times$ faster and reduces the setup payload by 31.6-83.0$\times$. Halevi-Shoup is unavailable at HE-32768 because its setup payload is too large to transmit over the network. Halevi-Shoup requires substantially more rotations and rotation keys, whereas the packed schedule of RowHyDiag reduces both costs.

\textbf{Comparison with the PEGASUS linear-transformation baseline.} Under the prototype settings, the two methods have comparable latency at HE-2048, although the PEGASUS baseline has a slightly smaller setup payload. At HE-4096, RowHyDiag is 1.1-1.7$\times$ faster and reduces the setup payload by 1.1-1.4$\times$. Under the configurations providing 128-bit security, RowHyDiag is 1.1-4.5$\times$ faster and reduces the setup payload by 1.2-3.4$\times$. These improvements arise from the RowHyDiag packing layout, which reduces both the number of diagonal products and the required rotation keys. For the larger CIFAR-100 and ImageNet-1K classifiers, BSGS scheduling preserves these advantages as the number of output classes increases.
}

\section{Conclusion}
\label{sec:conclusion}

This paper presents PRICE, a deployment-ready three-tier edge-cloud framework for privacy-preserving image classification with CKKS homomorphic encryption. The design keeps secret-key operations and deep feature extraction at the trusted edge while restricting the untrusted cloud to encrypted final-layer inference. To improve cloud-stage efficiency, we develop RowHyDiag, a row-packed hybrid diagonal encoding method with hoisted-rotation-aware computation, improving slot utilization and reducing costly homomorphic operations. \textcolor{blue}{Theoretical complexity analysis and empirical evaluation across five datasets demonstrate that, under the evaluated 128-bit-security configurations, RowHyDiag achieves 12.2-31.8$\times$ faster cloud inference, requires 32.0-85.3$\times$ fewer rotations, and reduces the setup payload by 31.6-83.0$\times$ compared with the Halevi-Shoup diagonal method. RowHyDiag is also 1.1-4.5$\times$ faster than the PEGASUS linear-transformation baseline and reduces its setup payload by 1.2-3.4$\times$. Encrypted accuracy matches plaintext accuracy across all five datasets, confirming that encrypted evaluation preserves the predictive results. These results show that our proposed framework supports low-latency encrypted inference for mobile edge-cloud systems under a trusted-edge threat model.}

\bibliographystyle{IEEEtran}
\bibliography{references}

\clearpage
\appendices

\section{Encoding Workflow Illustrations}
\label{app:encoding_illustrations}

\begin{figure}[t]
\centering
\includegraphics[width=0.48\textwidth]{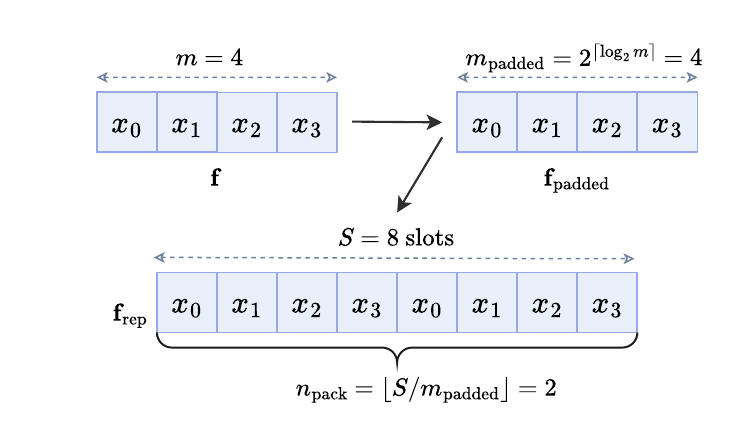}
\caption{Illustration of query encoding with power-of-two padding and slot replication. The feature vector $\mathbf{f}$ is padded to $\mathbf{f}_{\text{padded}}$ with $m_{\text{padded}}=2^{\lceil \log_2 m \rceil}$, then replicated across $S$ slots to form $\mathbf{f}_{\text{rep}}$ with $n_{\text{pack}}=\lfloor S/m_{\text{padded}} \rfloor$.}
\label{fig:query_encoding}
\end{figure}

\begin{figure*}[t]
\centering
\includegraphics[width=1\textwidth]{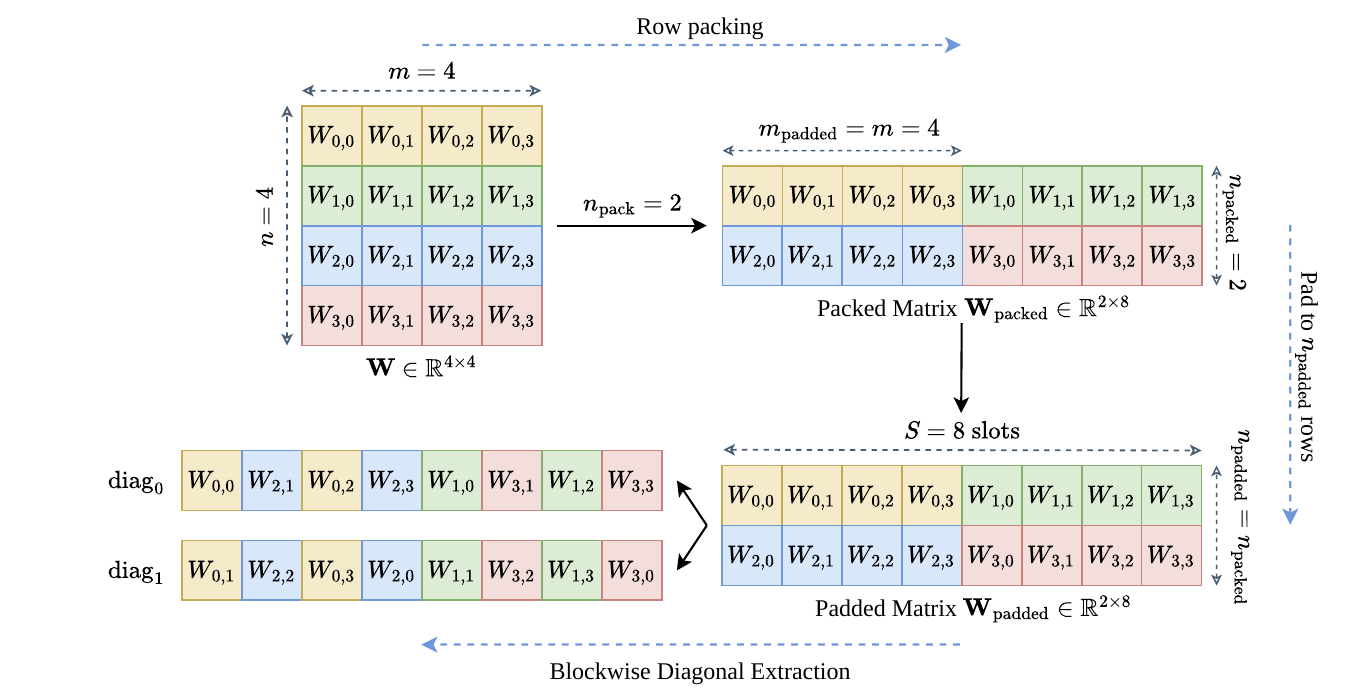}
\caption{Illustration of RowHyDiag matrix encoding. Row packing (square example): rows of $\mathbf{W}$ are packed to form $\mathbf{W}_{\text{packed}}$ with $m_{\text{padded}} = m$ and $n_{\text{packed}} = 2$. Pad to $n_{\text{padded}}$ rows: $\mathbf{W}_{\text{packed}}$ is aligned as the padded packed matrix. Blockwise diagonal extraction: diagonal plaintexts $\text{diag}_0$ and $\text{diag}_1$ are generated for encrypted linear-layer evaluation.}
\label{fig:hybrid_encoding_main}
\end{figure*}

\begin{figure}[t]
\centering
\includegraphics[width=0.50\textwidth]{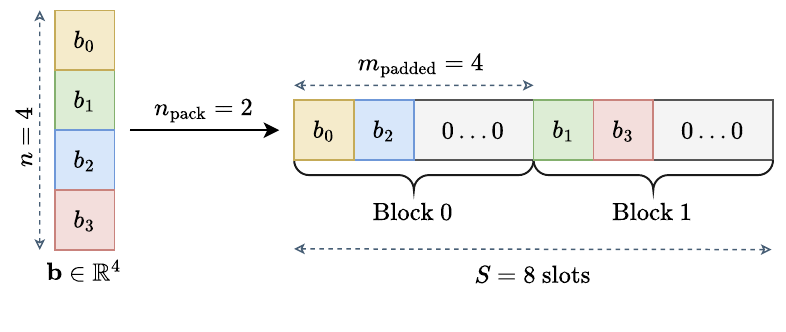}
\caption{Square-case bias encoding illustration ($4 \times 4$ example).}
\label{fig:bias_encoding_main}
\end{figure}

This appendix collects the main encoding workflow illustrations used to support the query encoding, RowHyDiag matrix encoding, and bias encoding procedures described in Section~III of the main manuscript, specifically Fig.~\ref{fig:query_encoding}, Fig.~\ref{fig:hybrid_encoding_main}, and Fig.~\ref{fig:bias_encoding_main}. These figures first show the square-case layouts, which provide the basic intuition for padding, slot replication, row packing, and bias placement.

\section{Non-Square Encoding Examples}
\label{app:nonsquare_examples}

Building on the square-case illustrations in \appref{app:encoding_illustrations}, this appendix provides non-square examples that clarify how the same RowHyDiag and bias-encoding rules apply when the number of rows and columns are not equal.

\begin{figure*}[t]
\centering
\includegraphics[width=0.82\textwidth]{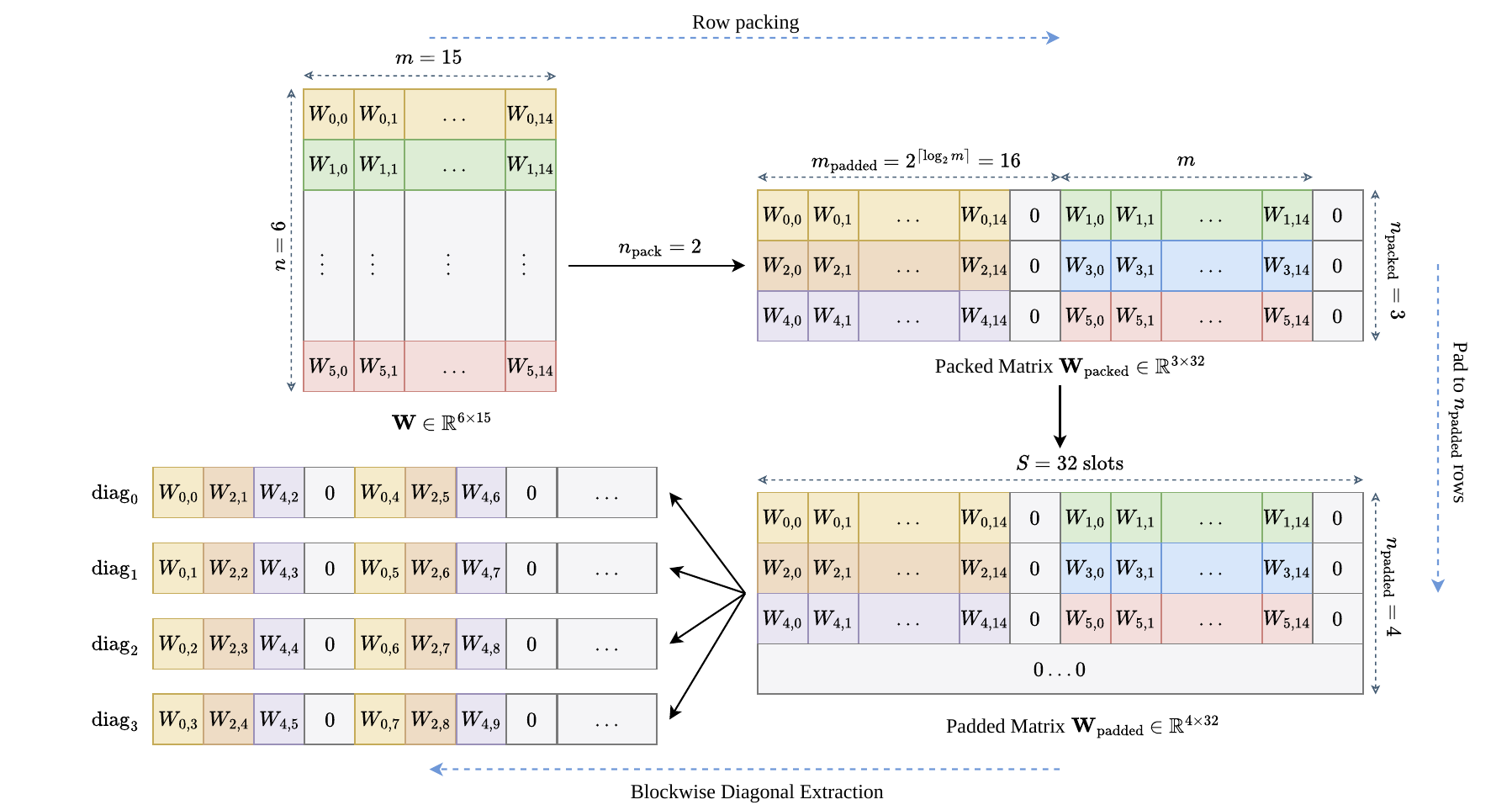}
\caption{Non-square-case workflow of RowHyDiag encoding ($6 \times 15$ example). Row packing: rows of $\mathbf{W}$ are packed to form a $3 \times 32$ matrix with $m_{\text{padded}}=16$ and $n_{\text{packed}}=3$ ($n_{\text{pack}}=2$). Pad to $n_{\text{padded}}$ rows: the packed matrix is zero-padded to $4 \times 32$ with $n_{\text{padded}}=4$. Blockwise diagonal extraction: diagonal plaintexts are generated from $\mathbf{W}_{\text{packed}}$ for encrypted matrix-vector computation.}
\label{fig:nonsquare_encoding_appendix}
\end{figure*}

For a representative non-square case, we consider an original weight matrix of size $6 \times 15$ under CKKS slot size $S=32$. Column alignment gives $m_{\text{padded}}=16$, and row packing with $n_{\text{pack}}=2$ produces a packed matrix of size $n_{\text{packed}} \times S = 3 \times 32$. To enable power-of-two scheduling in subsequent encrypted computation, the packed matrix is then padded to $n_{\text{padded}}=4$ rows, yielding $\mathbf{W}_{\text{packed}} \in \mathbb{R}^{4 \times 32}$. Fig.~\ref{fig:nonsquare_encoding_appendix} visualizes this full pipeline (row packing, row padding, and blockwise diagonal extraction) and clarifies how non-square dimensions map to the packed CKKS layout.

\begin{figure}[!htbp]
\centering
\includegraphics[width=0.50\textwidth]{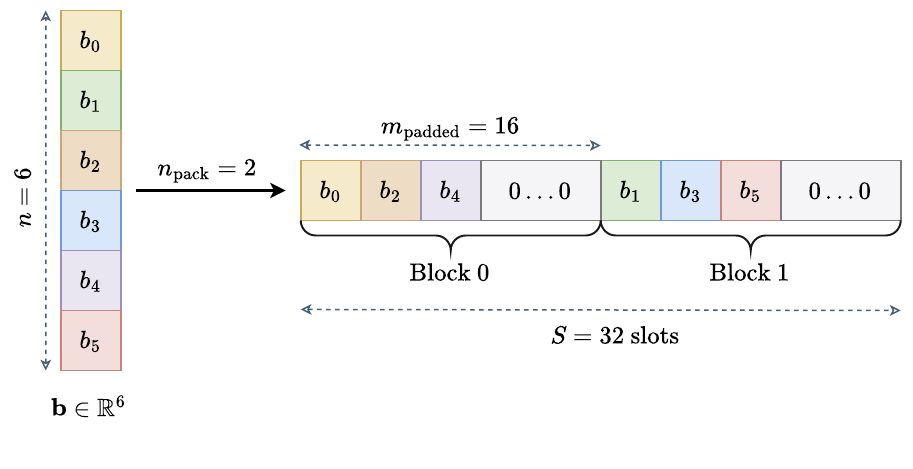}
\caption{Non-square bias encoding example. The original bias vector $\mathbf{b} \in \mathbb{R}^6$ is distributed across $n_{\text{pack}}=2$ blocks of size $m_{\text{padded}}=16$. Block 0 contains biases $b_0, b_2, b_4$ at positions 0, 1, 2, while block 1 contains biases $b_1, b_3, b_5$ at positions 16, 17, 18. Remaining positions are zero-padded.}
\label{fig:bias_encoding_appendix}
\end{figure}

For the same non-square setting ($n=6$, $m_{\text{padded}}=16$, $n_{\text{pack}}=2$), bias encoding follows the same block structure used by packed matrix-vector accumulation. The bias vector $\mathbf{b}\in\mathbb{R}^{6}$ is mapped by $i_{\text{bias}} = p\,n_{\text{pack}} + \text{blk}$, so valid class biases are placed at the leading positions of each block and all remaining slots are zero-filled. Fig.~\ref{fig:bias_encoding_appendix} illustrates this block-aligned placement, which ensures that ciphertext-plaintext bias addition remains index-consistent with the reduced accumulator slots.

\section{Automated Dataset Construction with Dual Verification}
\label{app:general_pipeline}

This appendix presents the general data-construction methodology used in this work, i.e., web crawling, semantic filtering, and visual verification. The goal is to describe a reusable pipeline design independent of any specific dataset.

To build a high-quality image dataset, we design a collection pipeline consisting of web crawling and dual-stage filtering. The pipeline ensures that collected images are diverse, visually relevant, and suitable for deep learning training. We first define target classes with distinguishable visual characteristics, then crawl publicly available images licensed for academic use, and finally apply two filtering stages to remove irrelevant or low-quality samples.

\textbf{Data Crawling}: Some prior works have crawled images from Flickr, Wikimedia Commons, and Google Images. A commonly used source is Flickr (e.g., Oxford Buildings and Paris datasets), where data are collected through the Flickr REST API using text queries, tags, and license filters. Wikimedia Commons is also widely used; for example, Google Landmarks Dataset v2~\cite{weyand2020google} is built from web-scale landmark sources. Because Flickr API access became restricted to Pro subscribers from August 2024, our framework uses Wikimedia Commons API category crawling. For each target class, we collect candidate images via category traversal and retain only Public Domain or research-permitted licensed content.

\textbf{Semantic Verification}: Crawled images may still include irrelevant content due to ambiguous names, incorrect tags, or noisy categories. We therefore use CLIP ViT-B/32~\cite{radford2021clip} for semantic filtering. For each class, we define multiple text prompts based on class names and characteristics, embed both text and images into CLIP space, compute similarities, and keep an image only if its maximum prompt similarity exceeds a predefined threshold.

\textbf{KNN-Based Visual Verification}: Even after semantic filtering, some images may remain visually mismatched (e.g., cartoons, misleading metadata, or unrelated scenes). We therefore apply a second stage based on visual similarity to high-quality reference galleries. For each candidate image, we compute similarity scores against gallery embeddings, retrieve the top-$k$ neighbors above threshold, and keep the image only if at least $\left\lceil k/2 \right\rceil$ of the $k$ nearest neighbors belong to the target class.

\section{Australian Landmarks Dataset Construction Statistics}
\label{app:landmarks_stats}

Building on the general pipeline in \appref{app:general_pipeline}, this appendix instantiates the same process for the 9-class Australian landmarks dataset and reports the final quantitative statistics used in the main experiments.

\subsection{Dataset Collection and Construction}
\label{app:dataset_collection}

To evaluate our privacy-preserving landmark recognition framework, we construct a dataset of 9 Australian landmarks using the two-stage verification pipeline. The primary crawler source is the Wikimedia Commons API, which provides public-domain and Creative Commons licensed images suitable for academic research.

We first identify Wikimedia category pages for each target landmark (e.g., \texttt{Category:Sydney\_Opera\_House}, \texttt{Category:Uluru}), then recursively traverse categories and subcategories to collect candidate images and metadata. For each image, we retain only samples with licenses permitting research use, enforce a minimum resolution of $224 \times 224$, and remove duplicates using MD5 hash matching.

The remaining candidates are filtered in two stages.
Stage 1 (CLIP semantic verification): with 9 landmarks and 4 prompts per landmark, we build 36 prompts and compute image-text similarity using CLIP~\cite{radford2021clip}. A sample is accepted only if its maximum similarity exceeds $\tau_{\text{CLIP}} = 0.24$ and the highest-scoring prompt belongs to the correct landmark class.
Stage 2 (reference-gallery visual verification): we run kNN retrieval against landmark-specific reference galleries selected for viewpoint and illumination diversity. A sample is accepted only if at least $\lceil k/2 \rceil$ of its $k=8$ nearest neighbors belong to the target class and exceed cosine similarity threshold $\tau_{\text{ref}} = 0.28$. Table~\ref{tab:filtering_stats} summarizes filtering statistics, and Table~\ref{tab:dataset_stats} reports the final train/test split. In Table~\ref{tab:filtering_stats}, the Stage 2 and Final columns are identical because no additional filtering is applied after Stage 2. The Australian Landmarks dataset is published on Hugging Face at \url{https://huggingface.co/datasets/huyn1602/Australian_Landmarks_Dataset}.

\begin{table}[t]
\centering
\caption{Dual-stage filtering pipeline statistics}
\label{tab:filtering_stats}
\begin{tabular}{lcccc}
\toprule
\textbf{Landmark} & \textbf{Crawled} & \textbf{Stage 1} & \textbf{Stage 2} & \textbf{Final} \\
 & & \textbf{Passed} & \textbf{Passed} & \textbf{Accepted} \\
\midrule
Sydney Opera House & 2,660 & 2,023 & 1,943 & 1,943 \\
Sydney Harbour Bridge & 2,522 & 1,959 & 1,836 & 1,836 \\
Uluru & 804 & 772 & 760 & 760 \\
Great Ocean Road & 1,601 & 1,345 & 1,277 & 1,277 \\
Great Barrier Reef & 2,112 & 1,762 & 1,700 & 1,700 \\
Blue Mountains & 780 & 570 & 521 & 521 \\
Flinders Street Station & 669 & 579 & 563 & 563 \\
Parliament House & 597 & 462 & 442 & 442 \\
Surfers Paradise & 1,156 & 973 & 842 & 842 \\
\midrule
\textbf{Total} & \textbf{12,901} & \textbf{10,445} & \textbf{9,884} & \textbf{9,884} \\
\midrule
\multicolumn{5}{l}{\textit{Filtering Effectiveness:}} \\
\multicolumn{2}{l}{Stage 1 pass rate:} & \multicolumn{3}{l}{80.96\% (CLIP semantic filter)} \\
\multicolumn{2}{l}{Stage 2 pass rate:} & \multicolumn{3}{l}{94.63\% (Reference visual filter)} \\
\multicolumn{2}{l}{Overall pass rate:} & \multicolumn{3}{l}{76.62\% (Dual-stage combined)} \\
\bottomrule
\end{tabular}
\end{table}

\begin{table}[t]
\centering
\caption{Final dataset statistics (train/test split)}
\label{tab:dataset_stats}
\begin{tabular}{lcc}
\toprule
\textbf{Landmark} & \textbf{Train (80\%)} & \textbf{Test (20\%)} \\
\midrule
Sydney Opera House & 1,554 & 389 \\
Sydney Harbour Bridge & 1,468 & 368 \\
Uluru & 608 & 152 \\
Great Ocean Road & 1,021 & 256 \\
Great Barrier Reef & 1,360 & 340 \\
Blue Mountains & 416 & 105 \\
Flinders Street Station & 450 & 113 \\
Parliament House & 353 & 89 \\
Surfers Paradise & 673 & 169 \\
\midrule
\textbf{Total} & \textbf{7,903} & \textbf{1,981} \\
\bottomrule
\end{tabular}
\end{table}

\section{Convergence Analysis of Model Fine-Tuning}
\label{app:convergence}

\begin{figure}[!t]
\centering

\begin{subfigure}[t]{\columnwidth}
\centering
\includegraphics[width=\linewidth]{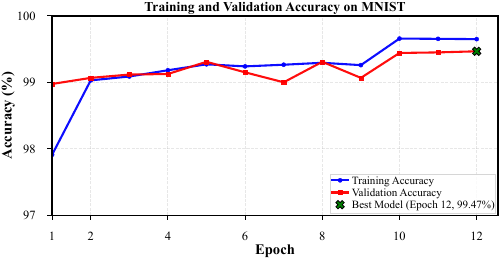}
\caption{}
\label{fig:conv_mnist}
\end{subfigure}

\begin{subfigure}[t]{\columnwidth}
\centering
\includegraphics[width=\linewidth]{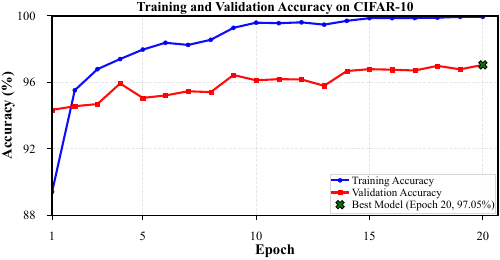}
\caption{}
\label{fig:conv_cifar10}
\end{subfigure}

\begin{subfigure}[t]{\columnwidth}
\centering
\includegraphics[width=\linewidth]{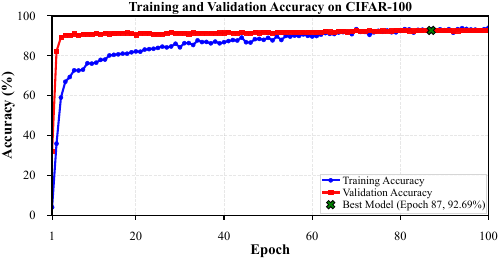}
\caption{}
\label{fig:conv_cifar100}
\end{subfigure}

\begin{subfigure}[t]{\columnwidth}
\centering
\includegraphics[width=\linewidth]{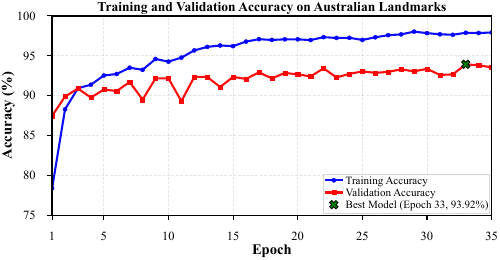}
\caption{}
\label{fig:conv_landmarks}
\end{subfigure}

\caption{Fine-tuning convergence across all evaluated datasets: (a) MNIST, (b) CIFAR-10, (c) CIFAR-100, and (d) Australian landmarks. ResNet-34 on MNIST, CIFAR-10, and Australian landmarks converges quickly with small train--validation gaps, while ViT-B/16 on CIFAR-100 converges more gradually but remains stable at a high-accuracy operating point.}
\label{fig:training_convergence_all}
\end{figure}

This appendix reports convergence behavior for all fine-tuned models to confirm that each reaches a stable operating point before encrypted inference. Fig.~\ref{fig:training_convergence_all} consolidates training and validation convergence across all four datasets and shows stable behavior without erratic validation fluctuations.

For MNIST (Fig.~\ref{fig:training_convergence_all}(a)), convergence is rapid: by epoch 12, training and validation accuracies reach 99.65\% and 99.47\%, respectively, with a very small gap. For CIFAR-10 (Fig.~\ref{fig:training_convergence_all}(b)), the model reaches 99.94\% training and 97.05\% validation accuracy at epoch 20; the larger but still controlled gap is consistent with higher natural-image variability and task difficulty compared with MNIST. For CIFAR-100 (Fig.~\ref{fig:training_convergence_all}(c)), ViT-B/16 converges more gradually under stronger augmentation and One-Cycle scheduling, with training accuracy increasing from 3.96\% to 94.06\% and validation accuracy from 32.15\% to 92.63\% over 100 epochs; the best validation point is 92.69\% at epoch 87.

For the Australian landmarks dataset (Fig.~\ref{fig:training_convergence_all}(d)), ResNet-34 improves from 78.38\% to 97.91\% training accuracy and from 87.47\% to 93.54\% validation accuracy across 35 epochs, with the best checkpoint at epoch 33 (93.92\% validation accuracy). Overall, these results confirm that all selected checkpoints (99.47\% MNIST, 97.05\% CIFAR-10, 92.69\% CIFAR-100, and 93.92\% landmarks) provide stable and reliable feature representations for encrypted final-layer inference. This is important because the encrypted stage evaluates only the final classifier on extracted embeddings; therefore, post-training convergence directly determines the performance upper bound of encrypted inference.

\end{document}